\documentclass[12pt]{article}
\usepackage{psfrag}
\usepackage{amsmath}
\usepackage{amsthm}
\usepackage{amssymb}
\usepackage{epsfig}
\usepackage{euscript}
\usepackage{array}
\usepackage{cite}
\usepackage{cancel}
\usepackage{mathtools}
\usepackage{empheq}
\usepackage{graphicx}
\usepackage{subfigure}
\usepackage{upgreek}
\usepackage{epstopdf}
\usepackage{booktabs}

\usepackage{hyperref}
\usepackage{verbatim}

\theoremstyle{definition}

\theoremstyle{remark}

\newcounter{multieqs}

\newcommand{\be}{\begin{equation}}
\newcommand{\ee}{\end{equation}}
\newcommand{\eq}[1]{(\ref{#1})}
\newcommand{\bit}{\begin{itemize}}  \newcommand{\eit}{\end{itemize}}
\newcommand{\ben}{\begin{enumerate}}  \newcommand{\een}{\end{enumerate}}

\newcommand{\bm}[1]{\mbox{\boldmath $#1$}}
\newcommand{\rf}[1]{(\ref{#1})}

\def\bd{\begin{document}}
\def\ed{\end{document}}
\def\bea{\begin{eqnarray}}
\def\eea{\end{eqnarray}}
\let\bm=\bibitem

\def\la{\langle}
\def\ra{\rangle}

\def\npb#1#2#3{Nucl. Phys. {\bf{B#1}} #3 (#2)}
\def\plb#1#2#3{Phys. Lett. {\bf{#1B}} #3 (#2)}
\def\prl#1#2#3{Phys. Rev. Lett. {\bf{#1}} #3 (#2)}
\def\prd#1#2#3{Phys. Rev. {D \bf{#1}} #3 (#2)}
\def\cmp#1#2#3{Comm. Math. Phys. {\bf{#1}} #3 (#2)}
\def\cqg#1#2#3{Class. Quantum Grav. {\bf{#1}} #3 (#2)}
\def\nppsa#1#2#3{Nucl. Phys. B (Proc. Suppl.) {\bf{#1A}}#3 (#2)}
\def\ap#1#2#3{Ann. of Phys. {\bf{#1}} #3 (#2)}
\def\ijmp#1#2#3{Int. J. Mod. Phys. {\bf{A#1}} #3 (#2)}
\def\rmp#1#2#3{Rev. Mod. Phys. {\bf{#1}} #3 (#2)}
\def\mpla#1#2#3{Mod. Phys. Lett. {\bf A#1} #3 (#2)}
\def\jhep#1#2#3{J. High Energy Phys. {\bf #1} #3 (#2)}
\def\atmp#1#2#3{Adv. Theor. Math. Phys. {\bf #1} #3 (#2)}

\def\N{{\cal N}}
\def\sst{\scriptscriptstyle}
\def\thetabar{\bar\theta}
\def\Tr{{\rm Tr}}
\def\one{\mbox{1 \kern-.59em {\rm l}}}

\def\a{\alpha}      \def\da{{\dot\alpha}}  \def\dA{{\dot A}}
\def\b{\beta}       \def\db{{\dot\beta}}  
\def\g{\gamma}  \def\G{\Gamma}  \def\dc{{\dot\gamma}}  
\def\d{\delta}  \def\D{\Delta}  \def\ddt{\dot\delta}  
\def\e{\epsilon}        
\def\ve{\varepsilon}  
\def\uve{\upvarepsilon}
\def\f{\phi}    \def\F{\Phi}    \def\vvf{\f}  
\def\h{\eta}  
\def\k{\kappa}  
\def\l{\lambda} \def\L{\Lambda}  
\def\m{\mu} \def\n{\nu}  
\def\o{\omega}  
\def\p{\pi} \def\P{\Pi}  
\def\r{\rho}  
\def\s{\sigma}  \def\S{\Sigma}  
\def\t{\tau}  
\def\th{\theta} \def\Th{\Theta} \def\vth{\vartheta}  
\def\X{\Xeta}  
\def\z{\zeta}  

\def\na{\nabla}  

\def\cA{{\cal A}} \def\cB{{\cal B}} \def\cC{{\cal C}}  
\def\cD{{\cal D}} \def\cE{{\cal E}} \def\cF{{\cal F}}  
\def\cG{{\cal G}} \def\cH{{\cal H}} \def\cI{{\cal I}}  
\def\cJ{{\cal J}} \def\cK{{\cal K}} \def\cL{{\cal L}}  
\def\cM{{\cal M}} \def\cN{{\cal N}} \def\cO{{\cal O}}  
\def\cP{{\cal P}} \def\cQ{{\cal Q}} \def\cR{{\cal R}}  
\def\cS{{\cal S}} \def\cT{{\cal T}} \def\cU{{\cal U}}  
\def\cV{{\cal V}} \def\cW{{\cal W}} \def\cX{{\cal X}}  
\def\cY{{\cal Y}} \def\cZ{{\cal Z}}

\def\ua{\underline{\alpha}}  
\def\uc{\underline{\phantom{\alpha}}\!\!\!\gamma}  
\def\um{\underline{\mu}}  
\def\ud{\underline\delta}  
\def\ue{\underline\epsilon}  
\def\una{\underline a}\def\unA{\underline A}  
\def\unb{\underline b}\def\unB{\underline B}  
\def\unc{\underline c}\def\unC{\underline C}  
\def\und{\underline d}\def\unD{\underline D}  
\def\une{\underline e}\def\unE{\underline E}  
\def\unf{\underline{\phantom{e}}\!\!\!\! f}\def\unF{\underline F}  
\def\unm{\underline m}\def\unM{\underline M}  
\def\unn{\underline n}\def\unN{\underline N}  
\def\unp{\underline{\phantom{a}}\!\!\! p}\def\unP{\underline P}  
\def\unq{\underline{\phantom{a}}\!\!\! q}  
\def\unQ{\underline{\phantom{A}}\!\!\!\! Q}  
\def\unH{\underline{H}}  
  
\def\As {{A \hspace{-6.4pt} \slash}\;}  
\def\bs {{b \hspace{-6.4pt} \slash}\;}  
\def\Ds {{D \hspace{-6.4pt} \slash}\;}
\def\Gts {{\Gt \hspace{-6.4pt} \slash}\;}
\def\ds {{\del \hspace{-6.4pt} \slash}\;}  
\def\ss {{\s \hspace{-6.4pt} \slash}\;}  
\def\ks {{ k \hspace{-6.4pt} \slash}\;}  
\def\ps {{p \hspace{-6.4pt} \slash}\;}   
\def\xs {{x \hspace{-6.4pt} \slash}\;}  
\def\pas {{{p_1} \hspace{-6.4pt} \slash}\;}  
\def\pbs {{{p_2} \hspace{-6.4pt} \slash}\;}   
\def\cFs {{{\cal F} \hspace{-6.4pt} \slash}\;}

\def\Ah{{\hat{A}}}  
\def\Dh{{\hat{D}}}
\def\Gh{{\hat{G}}}
\def\Fh{{\hat{F}}}
\def\Ih{{\hat{I}}} 
\def\Jh{{\hat{J}}} 
\def\Kh{{\hat{K}}}
\def\Lh{{\hat{L}}} 
\def\Ph{{\hat{P}}}
\def\Rh{{\hat{R}}}
\def\Vh{{\hat{V}}} 
\def\Xh{{\hat{X}}}
 
\def\ah{{\hat{\a}}}
\def\bh{{\hat{\b}}}
\def\gh{{\hat{\g}}}
\def\dh{{\hat{\d}}}
\def\hh{\hat{h}}
\def\uh{\hat{u}}  
\def\xh{\hat{x}}  
\def\yh{\hat{y}}  
\def\ph{\hat{p}}  
\def\xih{\hat{\xi}}  
\def\chih{\hat{\chi}}  
\def\Psih{\hat{\Psi}}    
\def\phih{\hat{\phi}}

\def\psit{\tilde{\psi}}  
\def\Psit{\tilde{\Psi}}   
\def\Psibt{\tilde{\bar{Psi}}}  

\def\st{\tilde{\sigma}}  

\def\delt{\tilde{\delta}}
\def\Phit{\tilde{\Phi}}   
\def\Phitb{\overline{\tilde{Phi}}}  
\def\tht{\tilde{\th}}  
\def\lt{\tilde{\l}}
\def\chit{\tilde{\chi}}   
\def\phit{\tilde{\phi}} 

\def\At{\tilde{A}}
\def\Bt{\tilde{B}}
\def\Ct{\tilde{C}}
\def\Dt{\tilde{D}}
\def\Et{\tilde{E}}
\def\Ft{\tilde{F}}
\def\Gt{\tilde{G}}
\def\Ht{\tilde{H}}
\def\It{\tilde{I}}
\def\Jt{\tilde{J}}
\def\Qt{\tilde{Q}}  
\def\Rt{\tilde{R}}  
\def\Mt{\tilde{M }}  
\def\Nt{\tilde{N}}   
\def\St{\tilde{S}}
\def\Vt{\tilde{V}}
\def\Xt{\tilde{X}} 
\def\at{\tilde{a}}
\def\ct{\tilde{c}}
\def\dt{\tilde{d}}
\def\htt{\tilde{h}} 
\def\ft{\tilde{f}}
\def\gt{\tilde{g}}
\def\pt{\tilde{p}}  
\def\qt{\tilde{q}}  
\def\vt{\tilde{v}}  
\def\nt{\tilde{n}}  
\def\ut{\tilde{u}}  
\def\wt{\tilde{w}}  
\def\zt{\tilde{z}} 
\def\xt{\tilde{x}} 
\def\yt{\tilde{y}} 
\def\Psit{\tilde{\Psi}}
\def\vphit{\tilde{\varphi}}  

\def\eb{\bar{\epsilon}} 
\def\delb{\bar{\partial}}  
\def\thb{\bar{\theta}}
\def\mub{\bar{\mu}}
\def\lamb{\bar{\l}}
\def\psib{\bar{\psi}}
\def\sb{\bar{\sigma}}
\def\xib{\bar{\xi}}
\def\chib{\bar{\chi}}

\def\Psib{\bar{\Psi}}
\def\Phib{\bar{\Phi}}
\def\Lamb{\bar{\Lambda}}
\def\Sb{{\overline \Sigma}}
\def\cb{\bar{c}}
\def\hb{\bar{h}}
\def\qb{\bar{q}}
\def\wb{\bar{w}}
\def\ub{\bar{u}}
\def\zb{{\bar{z}}}
\def\Hb{\bar{H}}
\def\Qb{{\bar Q}}
\def\Omegab{\overline{\Omega}}
\def\ob{\overline{\omega}}

\def\Ab{{\overline A}} \def\Bb{{\overline B}} \def\Cb{{\overline C}}  
\def\Db{{\overline D}} \def\Eb{{\overline E}} \def\Fb{{\overline F}}  
\def\Gb{{\overline G}} 
\def\Ib{{\overline I}}  
\def\Jb{{\overline J}} \def\Kb{{\overline K}} \def\Lb{{\overline L}}  
\def\Mb{{\overline M}} \def\Nb{{\overline N}} \def\Ob{{\overline O}}  
\def\Pb{{\overline P}}  \def\Rb{{\overline R}}  
 \def\Tb{{\overline T}} \def\Ub{{\overline U}}  
\def\Vb{{\overline V}} \def\Wb{{\overline W}} \def\Xb{{\overline X}}  
\def\Yb{{\overline Y}} \def\Zb{{\overline Z}}  

\def\fb{{\overline f}}
\def\gb{{\overline g}}
\def\mb{{\overline m}}
\def\lb{{\overline l}}
\def\yb{{\overline y}}
  
\def\ldel{{\overleftarrow{\del}}}
\def\rdel{{\overrightarrow{\del}}}
\def\ldeldel{{\overleftarrow{\del^2}}}
\def\rdeldel{{\overrightarrow{\del^2}}}
\def\ldelb{{\overleftarrow{\bar{\del}}}}
\def\rdelb{{\overrightarrow{\bar{\del}}}}

\def\ba{{\bf a}} 
\def\bk{{\bf k}}  
\def\bl{{\bf l}}  
\def\bp{{\bf p}}  
\def\bq{{\bf q}}  
\def\br{{\bf r}}
\def\bt{{\bf t}}
\def\bu{{\bf u}}
\def\bv{{\bf v}}
\def\bx{{\bf x}}  
\def\by{{\bf y}}  
\def\bR{{\bf R}}  
\def\bV{{\bf V}}

\def\bone{{\bf 1}}  

\def\va{{\vec a}}
\def\vk{{\vec k}}
\def\vp{{\vec p}}
\def\vq{{\vec q}}
\def\vx{{\vec x}}
\def\vy{{\vec y}}
\def\vu{{\vec u}}
\def\vv{{\vec v}}
\def \vH{{\vec H}}
\def \vg{{\vec g}}

\def\vs{{\vec \sigma}}
\def\vtau{{\vec \tau}}

\newcommand{\ov}[1]{\overrightarrow{#1}}

\def\frA{\mathfrak{A}}
\def\frB{\mathfrak{B}}
\def\frC{\mathfrak{C}}
\def\frD{\mathfrak{D}}
\def\frE{\mathfrak{E}}
\def\frF{\mathfrak{F}}
\def\frG{\mathfrak{G}}
\def\frH{\mathfrak{H}}
\def\frM{\mathfrak{M}}
\def\frN{\mathfrak{N}}
\def\frR{\mathfrak{R}}
\def\frW{\mathfrak{W}}

\def\fra{\mathfrak{a}}
\def\frb{\mathfrak{b}}
\def\frf{\mathfrak{f}}
\def\frg{\mathfrak{g}}
\def\frh{\mathfrak{h}}
\def\frl{\mathfrak{l}}
\def\frs{\mathfrak{s}}
\def\fri{\mathfrak{i}}
\def\frj{\mathfrak{j}}

\def\ma{\mathfrak{a}}
\def\mg{\mathfrak{g}}
\def\mh{\mathfrak{h}}
\def\mR{\mathfrak{R}}
\def\mN{\mathfrak{N}}

\def\d{\delta}\def\D{\Delta}\def\ddt{\dot\delta}  
  
\def\pa{\partial} \def\del{\partial}  
\def\xx{\times}  
\def\uno{\mbox{1 \kern-.59em {\rm l}}}    
  
\def\trp{^{\top}}  
\def\inv{^{-1}}  
\def\dag{{^{\dagger}}}  
\def\pr{^{\prime}}  
  
\def\rar{\rightarrow}  
\def\lar{\leftarrow}  
\def\lrar{\leftrightarrow}  
  
\newcommand{\0}{\,\!}      
\def\one{1\!\!1\,\,}  
\def\im{\imath}  
\def\jm{\jmath}  
  
\newcommand{\tr}{\mbox{tr}}  
\newcommand{\slsh}[1]{/ \!\!\!\! #1}  
  
\def\vac{|0\rangle}  
\def\lvac{\langle 0|}  
  
\def\hlf{\frac{1}{2}}  
\def\ove#1{\frac{1}{#1}}  

\def\Box{\square}  
\def\CC {\mathbb{C}}
\def\FF {\mathbb{F}}
\def\RR{\mathbb{R}}
\def\NN{\mathbb{N}}  
\def\ZZ{\mathbb{Z}}  
\def\bb#1{{\bf #1}}  
\def\bcomment#1{}  
\def\bfhat#1{{\bf \hat{#1}}}  
\def\VEV#1{\left\langle #1\right\rangle}  

\newcommand{\ex}[1]{{\rm e}^{#1}} \def\ii{{\rm i}}  

\newcommand{\lrbrk}[1]{\left(#1\right)}
\newcommand{\lrsbrk}[1]{\left[#1\right]}
\newcommand{\sfrac}[2]{{\textstyle\frac{#1}{#2}}}
 
\def\stw{{\sqrt{2}}}

\def\rf {{\rm f}}
\def\ri {{\rm i}}
\def\rj {{\rm j}}
\def\rn {{\rm n}}
\def\rk {{\rm k}}
\def\rl {{\rm l}}
\def\rs {{\scriptscriptstyle \rm S}}
\def\rt {{\scriptscriptstyle \rm T}}

\def\rQ {{\scriptscriptstyle \rm \cQ}}
\def\rR {{\scriptscriptstyle \rm \cR}}

\def\cQb{{\cal \Qb}}
\def\cRb{{\cal \Rb}}
\def\cWb{{\cal \Wb}}

\def\fd {{\rm N}}
\def\afd {{\overline{\rm N}}}

\def \II {I\hspace{-.1em}I\hspace{.1em}}
\def \IIA {\mbox{\II A\hspace{.2em}}}
\def \IIB {\mbox{\II B\hspace{.2em}}}
\def \gs {g^s}
\def \ls {\lambda^s}

\def \I {{\cal I}}
\def \qs {q\hspace{-.53em}/\hspace{.15em}}
\def \ks {k\hspace{-.53em}/\hspace{.15em}}
\def \YM {{\mbox{\tiny YM}}}
\def \gym {g_{\YM}}

\def \Lc {\L_c}
\def\IR{\relax{\rm I\kern-.18em R}}
\def \id {{\bf 1}}

\def\cci{\ell}
\def\ccj{\ell'}

\def \thbb{\overline{\th\th}}
\newcommand \ol{\overline}
\def \lamb{\bar{\lambda}}
\def \vphi{\varphi}
\def \lambh{\hat{\bar{\lambda}}}
\def \lh{\hat{\lambda}}
\def \dd{\ddagger}

\newcommand{\QNB}[3]{[#1,#2,#3]}
\def\hm{\tilde{\eta}} 
\def\lp{l_{+}}
\def\lm{l_{-}}
\def \PS {{(\text{PS})}}
\def \Dir {{(\text{Dirac})}}
\def \WY {{(\text{WY})}}
\def \Sin {{(\text{Sin})}}
\def \tHP{{(\text{'t-P})}}
\def \uo {{U(1)}}
\def \Lt {\tilde{L}}
\def \tn {{\tau}^{(n)}}
\def \thn {{\hat{\theta}}^{(n)}}
\def \vphin {{\hat{\vphi}}^{(n)}}

\newcommand{\Ga}{{\Gamma}}
\newcommand{\De}{{\Delta}}
\newcommand{\Lm}{{\Lambda}}
\newcommand{\Om}{{\Omega}}

\newcommand{\al}{{\alpha}}
\newcommand{\ga}{{\gamma}}
\newcommand{\de}{{\delta}}
\newcommand{\ep}{{\epsilon}}
\newcommand{\vep}{{\varepsilon}}
\newcommand{\te}{{\theta}}
\newcommand{\ka}{{\kappa}}
\newcommand{\vpi}{{\varpi}}
\newcommand{\sig}{{\sigma}}
\newcommand{\om}{{\omega}}

\newcommand{\alt}{{\rm alt}}
\newcommand{\bdy}{{\rm bdy}}
\newcommand{\bsa}{{\boldsymbol{a}}}
\newcommand{\bsb}{{\boldsymbol{b}}}
\newcommand{\bsD}{{\boldsymbol{D}}}
\newcommand{\bsk}{{\boldsymbol{k}}}
\newcommand{\bsM}{{\boldsymbol{M}}}
\newcommand{\bulk}{{\rm bulk}}
\newcommand{\cont}{{\rm cont.}}
\newcommand{\cdN}{{\mathcal{N}_d}}
\newcommand{\lan}{{\langle}}
\newcommand{\pd}{{\partial}}
\newcommand{\R}{{\rm R}}
\newcommand{\rad}{{\rm rad}}
\newcommand{\ran}{{\rangle}}
\newcommand{\Slash}[1]{{\ooalign{\hfil/\hfil\crcr$#1$}}} 
\newcommand{\srel}[2]{{\stackrel{\scriptstyle #1}{\scriptstyle #2}}}
\newcommand{\std}{{\rm std}}
\newcommand{\U}{{\rm U}} 
\newcommand{\ul}{\underline}
\newcommand{\UV}{{\rm UV}}
\newcommand{\wg}{{\wedge}}
\newcommand{\wh}{\widehat}

\def\Log{\mathop{\rm Log}}
\def\Spin{\mathop{\rm Spin}}
\def\SO{\mathop{\rm SO}}
\def\O{\mathop{\rm O}}
\def\SU{\mathop{\rm SU}}
\def\U{\mathop{\rm U}}
\def\Sp{\mathop{\rm Sp}}
\def\SL{\mathop{\rm SL}}
\def\GL{\mathop{\rm GL}}

\def\det{\mathop{\rm det}\nolimits}
\def\sign{\mathop{\rm sign}\nolimits}
\def\mod{\mathop{\rm mod}\nolimits}
\def\tr{\mathop{\rm tr}\nolimits}
\def\diag{\mathop{\rm diag}\nolimits}
\def\Re{\mathop{\rm Re}\nolimits}
\def\Im{\mathop{\rm Im}\nolimits}
\def\Tr{\mathop{\rm Tr}\nolimits}
\def\bbra{{\langle\kern-2.5pt\langle}}
\def\kket{{\rangle\kern-2.5pt\rangle}}
\def\Bbra{{\Big\langle\kern-3.5pt\Big\langle}}
\def\Kket{{\Big\rangle\kern-3.5pt\Big\rangle}}

\newcommand{\nn}{\nonumber}
\newcommand{\Ord}{{\cal O}}
\def\a{\alpha}
\def\amm{\alpha^{\mu}_m}
\def\amM{\alpha^{\mu}_{-m}}
\def\anm{\alpha^{\nu}_m}
\def\anM{\alpha^{\nu}_{-m}}
\def\amn{\alpha^{\mu}_n}
\def\amN{\alpha^{\mu}_{-n}}
\def\ann{\alpha^{\nu}_n}
\def\anN{\alpha^{\nu}_{-n}}
\def\b{\beta}
\def\Del{\Delta}
\def\ep{\epsilon}
\def\vep{\varepsilon}
\def\ome{\omega}
\def\Ome{\Omega}
\def\lam{\lambda}
\def\Lam{\Lambda}
\def\m{\mu}
\def\n{\nu}
\def\t{\tau}
\def\sig{\sigma}
\def\parl{\partial}
\def\bigtri{\bigtriangleup}
\def\tri{\triangle}
\def\gam{\gamma}
\def\r{\rho}

\newcommand {\beq}{\begin{eqnarray}}
\newcommand {\eeq}{\end{eqnarray}}

\newcommand\blue[1]{{\textcolor{blue}{#1}}} 
\newcommand\red[1]{{\textcolor{red}{#1}}} 
\newcommand\uline[1]{{\underline{#1}}}

\renewcommand{\theequation}{\thesection.\arabic{equation}}
\makeatletter
\renewcommand{\theequation}{%
\thesection.\arabic{equation}}
\@addtoreset{equation}{section}
\makeatother

\begin{document}
\begin{titlepage}

\begin{flushright}
\hfill{NCTS-TH/1501}\\  
\today
 \end{flushright}
\hfill 

\begin{center}
 
{\Large \bf Dilaton, \\
Screening of the Cosmological Constant 
and IR-Driven Inflation 
}\\[10mm] 

{\bf Chong-Sun Chu${\,}^{a,b}$ and Yoji Koyama${\,}^{a}$}

${}^a\,${\itshape Physics Division, National Center for Theoretical Sciences, \\
 National Tsing-Hua University, Hsinchu, 30013, Taiwan}\\[1mm]
${}^b\,${\itshape Department of Physics, National Tsing-Hua
  University,  Hsinchu 30013, Taiwan}\\

\end{center}
\begin{abstract}
It is known that infrared (IR) quantum fluctuations in de Sitter space
could break the de Sitter symmetry and generate time dependent
observable effects. In this paper, we consider a dilaton-gravity
theory. We find that gravitational IR effects lead to a time dependent
shift on the vev of the dilaton and results in a screening (temporal)
of
the cosmological constant/Hubble parameter.  In the Einstein frame,
the effect is exponentiated and can give rises to a much more notable amount of
screening.  Taking
the dilaton as inflaton,  we obtain an inflationary expansion of the
slow roll kind. This inflation is
driven by the IR quantum effects of de Sitter gravity and does not rely
on the use of a slow roll potential.  As a result, our model is free from the eta
problem which baffle the standard slow roll inflation models.

\end{abstract}

\end{titlepage}

\newpage
\tableofcontents

\newpage

\section{Introduction}

It is  widely believed that the universe has
underwent a period of accelerated expansion in the early
cosmology. Such a period of inflation not only solves  
the flatness and horizon problems of the standard cosmology, but also,
with the introduction of an inflaton scalar field and an almost flat
potential,
predicts a nearly scale invariant density perturbation.
This picture is in excellent agreement with observational results of
the cosmic microwave background (CMB) and the large scale structure of
the universe. However this simple
picture is not without problem as it has proven 
extremely difficult to bring together the inflationary paradigm 
with fundamental particle physics.
For example, Planck mass suppressed
corrections to the inflaton potential generally
lead to large corrections to the
inflaton mass, resulting in a large
slow roll parameter $\eta$ which renders prolonged slow roll inflation
impossible. One may resort to symmetries such as supersymmetry,
global symmetries or higher dimensional gauge symmetries 
to protect the potential. However supersymmetry only
alleviates the problem as supersymmetry must be spontaneously broken at
the Hubble scale during inflation; while in the latter approaches
approach, a precise control of the Planck suppressed operators breaking
the symmetry is needed, and hence the necessity of a full treatment in a theory 
quantum gravity, e.g. string theory. Interesting effective theories
with novel physical effects have been inspired and constructed in string theory, for
example, D-brane
inflation \cite{kklt}, DBI inflation \cite{dbi} and axion
monodromy inflation \cite{axion}.  
Nevertheless, the construction of inflation model with controllable quantum
corrections remains a significant obstacle.

One of the motivation of this work is
the desire to come up with a new mechanism to drive inflation that
does not employ a slow roll potential. 

Dark energy presents another deep mystery of the
universe. A common feature shared by both 
inflation and dark energy 
is that both involve a
cosmological expansion described by a de Sitter
metric. 
The current cosmological constant is of the order of $10^{-120}
M_P^2$. The deep mystery of the cosmological constant problem
\cite{ccp} is to
understand why is there such a huge hierarchy of scales between the
current cosmological constant and the Planck scale 
or some high energy scale. 
It is natural to suspect that a
good understanding of the quantum properties of de Sitter space would
be necessary in order to tackle this problem properly.  
It has been conjectured some time ago
that IR
quantum effects in de Sitter space could lead to a kind of screening
to the cosmological constant \cite{poly,ford} and provide a resolution
to the cosmological constant problem. 
Explicit de Sitter breaking IR effect has been identified
\cite{VF,Chernikov:1968zm,Allen:1987tz,Allen:1986dd} 
and demonstrated to lead to a 
weakening effect (over time) on the
cosmological constant \cite{woodard96}. Similar effects on the
couplings of the matter sector have also been found 
\cite{Kitamoto:2012ep,Kitamoto:2014gva}. 
However the effects of the
screening as obtained from the perturbation theory are usually small. 
This is still the case even if one may re-sum the perturbation
result and extend it's regime of validity. For other studies of IR
effects of quantum theory
in de Sitter space, see for example \cite{Kleppe:1993fz}-\cite{Tanaka:2014ina}.

Another
motivation of this work is to identify new concrete mechanism for the
screening of the cosmological constant in which the screening can be
sizable, and yet the theory remains in a reliable regime.

In this paper, we consider a theory consisting
of a dilaton field coupled to gravity. This is part of 
the low energy effective theory of the NS-NS sector in any string model building.
The dilaton has a potential but
its detailed form is not important to us. We will only need to assume
that the theory admits a minimum where the dilaton is taking a
vacuum expectation value (vev)  $\phi = v, \quad v \in \bR$, 
and the corresponding potential energy is positive;  thereby 
giving rises to a de Sitter metric. As is well known for de Sitter
space, the quantum fluctuations of 
a massless minimal coupled scalar field grows linearly with time
and breaks the de Sitter symmetry 
\cite{VF}. In fact the propagator 
of a massless minimal coupled scalar is IR
divergent (see \eq{mmc} below) and breaks the de Sitter symmetry
explicitly \cite{Chernikov:1968zm,Allen:1987tz,Allen:1986dd}. 
The time dependent origin of the IR divergence is simple 
and can be traced back to the  exponential increase in
the number of degrees of freedom outside the 
Hubble horizon. The dilaton in our theory is massive and is not minimally
coupled. However part of the graviton excitation modes 
are massless minimally coupled and so the time dependent IR
effects inherent in them could generate time dependent effects on other
physical quantities of theory. In this paper, we 
identify such an effect on the vev of the dilaton field. We find
that the IR effects of the gravitational loops (one loop order)
induce a time dependence in the vev 
\be \label{vt}
v \to v_{\rm eff} := v + \d v(t), 
\quad
\mbox{where $\d v \propto  (H_0 t)^2$}.
\ee
Here $H_0$ is the Hubble parameter.

As the modification grows quadratically with time, perturbation theory
will eventually break down and this imposes a serve limit on the size
of this effect. 
In particle theory, it is
possible to re-sum the leading order time dependent corrections 
by employing the dynamical renormalization group (DRG) equation 
\cite{vega}
and obtain a result which has a much bigger regime of validity. 
The  resummation of the IR divergences and the understanding of the
associated  late time secular evolution in quantized gravity is,
however,  a much more difficult  open
problem. For the case of a massive scalar field in de Sitter space with
non-derivative self-interaction, the problem is simpler. 
Over the years various approaches have been proposed and considered,
 for example, the  
semiclassical stochastic methods \cite{sto}, DRG \cite{burgess, bur1},  Schwinger-Dyson equation
\cite{SD}. Although qualitatively similar results are obtained, e.g. on
the generation of dynamical mass \cite{Garbrecht-self}, these different
approaches do not report exactly the same results on the 
non-perturbative resummation. This is presumably due to 
different aspect of physics were being emphasized and hence slightly 
different approximations and assumptions have been made
correspondingly. For example,  stochastic inflation relies on the
assumption of a Gaussian probability distribution for the background
quantities. In the approaches of Schwinger-Dyson or DRG, it is inevitable to truncate the full set of
Feynman diagrams or renormalization group (RG) equations to a manageable subset, leading to
disregarding set of diagrams or RG flows that may not be subleading at
all in the IR \cite{riotto}.
In  general model with derivative interaction, including the
case of gravity, the problem is much more difficult and much less is
known: the Schwinger-Dyson
equation is far too complicated; and it is not know
how to generalize the stochastic approach in this case (see however
\cite{woodard-inf} 
for
some suggestions). Comparatively,  the DRG approach is relatively simpler. Therefore 
in this paper we will adopt the DRG approach to resum the leading IR divergences.
We believe the time
dependent behaviour we found are qualitatively correct although the
precise details may be different.
We do
warn the reader that we are not claiming that we have solved the
important open problem of determining the secular IR effects of quantized gravity. 

Due to the 
time dependence \eq{vt} of the vev, the Hubble parameter of the theory
becomes time dependent and decreases slowly with time. That it is
slowly changing can be confirmed from the small values of the associated 
slow roll parameters. 
Since these slow roll parameters measure the back reaction of the quantum
effects on the classical de Sitter background, their smallness means  we can
trust the result of the perturbative computation. 

The above analysis are performed in the string frame. To examine the
physical significance of the time dependence of the vev, we need to go
to the Einstein frame.  As a result of the change of frame, 
the time dependent effect gets magnified exponentially. We have
thus obtained a mechanism of screening of the cosmological constant 
where a significant amount of screening can be achieved within a
calculable and reliable framework. The screening is 
due to the de Sitter symmetry breaking IR effect of the graviton
loops.  The behavior of the Hubble
parameter in the Einstein frame is one of the slow roll inflation. However
there is no almost flat potential and inflation is not achieved by the 
slow rolling of the inflaton field.   Therefore
we have obtained a model of inflation where the inflationary expansion
is driven by the de Sitter symmetry breaking gravitational 
IR effect. In particular, it
is important to note that since our mechanism does not rely on the
existence of a slow roll potential, our model is free from the eta
problem which baffles the standard slow roll inflation models.

The plan of the paper is as follows. In section 2, we introduce
our model. The classical solution of the theory is discussed in
section 2.1. In section 2.2, we set up the perturbation theory. In section
2.3, we use the in-in formalism to compute the time dependent IR
corrections 
on the vev at 1 loop order. In section 2.4, we compute the slow roll
parameters and demonstrated that they are small.
We then go to the Einstein frame in section 3. In section 3.1, 
by demanding that the Planck mass is time
independent, we fix
the choice of frame and obtain the Einstein frame Hubble parameter.
This is shown to be of  slow roll type in section 3.2. In section 3.3,
we look at some different choices of parameters 
of the model 
and demonstrate that
the de Sitter symmetry breaking IR effect could provide sufficient
screening during inflation. This offer an explanation of 
why the Hubble parameter during
inflation is so much smaller compared to the Planck scale  
(approximately $10^{-4}$ of it).

\section{Perturbative Analysis in the String Frame}

We consider Einstein gravity coupled non-minimally
to a scalar field $\phi$ described by
the action
\be
S=\int\sqrt{-{ g}}d^4 x\left[\frac{M^2}{2}e^{-2\phi/\eta}{
    R}- \frac12{ g}^{\mu\nu}\parl_{\mu}\phi\parl_{\nu}\phi-V(\phi)\right].
\label{action}\\
\ee 
Here $M$ is a mass parameter that set the fundamental scale of the theory.  
The coupling of the scalar field
$\phi$ to gravity is described by an exponential coupling
$e^{-2\phi/\eta}$
whose strength is controlled by the the dimensionful
coupling constant $\eta$. The limit $\eta\to \infty$ 
corresponds to a minimally coupled theory, 
where
the effects we computed in this paper, e.g. \eq{vtot-1}, will go away.
We have also included a potential term $V$. We will not need to assume
any particular details for it except for the assumption
that a vev $v>0$ is developed for a stable vacuum.

The action \eq{action} without the potential term is simply 
the Jordan-Brans-Dicke
theory \cite{JBD, JBD-g},
in which case the scalar
field is simply a phenomenological possibility.  In string theory,
$\phi$ arises necessary as the dilaton of the closed string
sector. 
The general form of the string frame, gravi-dilaton low energy 
effective action, to the lowest order in the $\a'$,
has been argued to take the form \cite{ven},
\be \label{action1}
S=\frac{M^2}{2} \int\sqrt{-{ g}}d^4 x\left[ B_g(\phi) R- \frac{B_\phi(\phi)}{2}
  (\del \phi)^2 + \frac{2}{M^2} V(\phi) \right].
\ee
Here the ``form factors'' $B_g (\phi), B_\phi(\phi)$ include 
the dilaton-dependent loop corrections and other effects of the
background flux, compactification, branes configuration etc; and 
$V(\phi)$ is an effective dilaton potential.   
The action \eq{action}  can be
considered as a  special case  of \eq{action1}. 
In this paper we will take the  phenomenological approach without
worrying the embedding of our action in string theory.
Our model is specified by the value of the parameter $\eta$ 
and the potential $V$. However our results do not depend on 
the specific knowledge of $V$ 
\footnote{
Incidentally the kind of action \eq{action} has also been considered
recently by \cite{lin} as a proposal to solve the gauge hierarchy problem. 
In this paper, specific details of the potential for spontaneous symmetry breaking is needed, 
e.g. a certain large value of the classical vacuum expectation value
of the dilaton is suggested to give rises to the hierarchy.  
}.

\subsection{The de-Sitter vacuum}

We begin by analyzing the classical dynamics of the theory
\eq{action}. The equations of motion are:
\be \label{seq} 
\Box
\phi-V'(\phi)-\frac{M^2}{\eta}e^{-2\phi/\eta}{ R}=0,
\ee
\be 
R_{\mu\nu}-\frac12{ R}{ g}_{\mu\nu}=\frac{1}{M^2}{\bar
  T}_{\mu\nu}.\label{geq} 
\ee 
Here $ \Box \phi=\frac{1}{\sqrt{-{ g}}}\partial_{\mu}(\sqrt{-{ g}} 
{g}^{\mu\nu}\partial_{\nu}\phi)$ and 
\be {\bar
  T}_{\mu\nu} =e^{2\phi/\eta}\left[\parl_{\mu}\phi\parl_{\nu}\phi-
  g_{\mu\nu}
(\frac{1}{2}(\pa \phi)^2 + V(\phi)) +M^2(D_{\mu}D_{\nu}-g_{\mu\nu}
  D^2)
e^{-2\phi/\eta} \right], 
\ee 
and $D_{\mu}$ is the covariant
derivative. The scalar equation can be simplified by substituting the
trace of the Einstein equation \eqref{geq} 
\be { R} =
\frac{4}{M^2}e^{2\phi/\eta}V(\phi)
+\frac{e^{2\phi/\eta}}{M^2}\left(1+12\frac{M^2}{\eta^2}e^{-2\phi/\eta}\right)
\parl_{\mu}\phi\parl^{\mu}\phi
-
\frac{6}{\eta}\Box
\phi \label{riccitrace} 
\ee 
into (\ref{seq}) and gets
\be \label{seq2}
\Box
\phi-\left(1+6\frac{M^2}{\eta^2}e^{-2\phi/\eta}\right)^{-1}
\left(
V'(\phi)+\frac{4}{\eta}V(\phi)
+\frac{1}{\eta}
(1+12\frac{M^2}{\eta^2}e^{-2\phi/\eta})\parl_{\mu}\phi\parl^{\mu}\phi
\right)=0. \quad
\ee
A solution is given by the  de Sitter background: 
\beq
&&\phi=v \ (\text{const.}), \quad
\mbox{such that} \quad V'(v)+\frac{4}{\eta}V(v)=0, \label{seq3} \\
&& R^{(0)}=\frac{4}{M^2}e^{2v/\eta}V(v) . \label{dscur}
\eeq
We note that  the field equation \eq{seq3} for a constant scalar $\phi$
configuration is modified from the usual one by the second
term due to its non-minimal coupling to gravity. To get a de-Sitter
background, we assume that
\be \label{V1}
V(v) >0.
\ee

Let us also investigate the local stability of the de Sitter solution
against small variations. 
Consider
\beq
\phi&=&v+\delta \phi,\\
{ g}_{\mu\nu}&=&{ g}^{\rm dS}_{\mu\nu}+\delta{ g}_{\mu\nu} \rightarrow
{ R}
=R^{(0)}+\delta{ R}.
\eeq
At ${\cal O}(\delta\phi,\delta{ R})$, the scalar field equation becomes
\beq
\Box\delta\phi+(-V''(v)+\frac{2M^2}{\eta^2}e^{-2v/\eta}R^{(0)})\delta\phi
-\frac{M^2}{\eta}e^{-2v/\eta}\delta{ R}=0.\label{sfleq}
\eeq
On the other hand, $\delta R$ satisfies
\beq
\delta{
  R}=\frac{2}{\eta}R^{(0)}\delta\phi+\frac{4}{M^2}e^{2v/\eta}V'(v)\delta\phi 
-\frac{6}{\eta}\Box \delta\phi.\label{rfleq}
\eeq
Put $\delta{ R}$ into (\ref{sfleq}) we obtain 
\beq
\Box\delta\phi-\left(1+\frac{6M^2}{\eta^2}e^{-2v/\eta}\right)^{-1}
\left(V''(v)+\frac{4}{\eta}V'(v)\right)\delta\phi=0
\eeq
The solution $\phi=v$ is stable for 
\beq
V''(v)+\frac{4}{\eta}V'(v)>0.\label{lsta}
\eeq
The conditions \eq{V1}, \eq{seq3}, \eq{lsta} are all that we will assume
for the potential $V(\phi)$. This can be easily satisfied. For
example any potential that behaves near $\phi =v$ as
\be
V(\phi) = V_0 + \rho (\phi -v) + \frac{\s^2}{2} (\phi -v)^2
\ee
with 
\be
V_0:=V(v) >0, 
\ee
\be 
\rho := V'(v)= -\frac{4 V_0}{\eta}, 
\ee
and
\be\label{sigmastable}
\s^2:=V''(v) > \frac{16 V_0}{\eta^2} 
\ee
are all right.

\subsection{Scalar and graviton propagators}

Our central interest is to calculate the graviton one-loop corrections to
the vev of the dilaton scalar field in the de Sitter space.  We consider the
Poincar{\'e} coordinate
\be \label{g0-string}
ds^2 = -dt^2 + e^{2H_0t} d\bx^2 = a_0(\t)^2 (- d\t^2 + d\bx^2), 
\ee
where the Hubble constant is given by the Friedmann equation
\be \label{H0-tree}
H_0^2=\frac{1}{3 M^2}e^{2v/\eta}V(v),
\ee
and $\t$ is the conformal time  
\be
\tau=-\frac{1}{H_0}e^{-H_0t}.
\ee
The scale factor is 
\be
a_0(\tau)=-\frac{1}{H_0\tau}. 
\ee  
In this paper we will use $H_0$ and $H$ to denote the string frame 
Hubble constant at tree level and at 1-loop, and 
 $H_{E0}$ and $H_E$ to denote the Einstein frame 
Hubble constant at tree level and at 1-loop.

To obtain the perturbative action, we expand $\phi$  around its
classical value, 
\be \phi=v+{\tilde \phi}.  
\ee
In order to
construct the scalar and graviton propagators, we begin with the
quadratic action in the perturbations:
\beq
S^{(2)}&=&\int d^4 x
\left[
\frac{1}{\kappa^{2}}
\Big((\sqrt{-{g}}{R})^{(2)}-6(\sqrt{-{g}})^{(2)}H_0^2\Big)+
\Big(\frac{2}{\kappa^2\eta}(\sqrt{-{g}}{R})^{(1)}-(\sqrt{-{g}})^{(1)}\r\Big)
     {\tilde \phi} \right.\notag\\ 
&&\qquad \qquad
     \left. -\frac12 a_0^2\eta^{\mu\nu}\parl_{\mu}{\tilde
       \phi}\parl_{\nu}{\tilde \phi}
     +\frac12 a_0^4\left(\frac{48}{\kappa^2\eta^2}H_0^2+\sig^2\right){\tilde
       \phi}^2 \right].
\label{saction} 
\eeq 
We shall adopt the same parametrization of the graviton perturbations
as \cite{Kitamoto:2012ep,Kitamoto:2014gva},
\beq
&&{g}_{\mu\nu}=\Omega^2(x){ {\tilde g}}_{\mu\nu},\quad 
\det {{\tilde g}}_{\mu\nu}=-1\notag\\
&&\Omega(x)=a_0(\tau)e^{\kappa \omega(x)},\quad 
{ {\tilde g}}_{\mu\nu}=\left(e^{\kappa h}\right)_{\mu\nu},
\quad \kappa^2=\frac{M^2}{2}e^{-2\frac{v}{\eta}},
\label{kkh}
\eeq
where $\omega$ is the perturbation of the conformal mode, 
$h$ is the traceless tensor perturbation. 
Indices on $h_{\mu\nu}$ are raised and lowered with the 
Lorentz metric $\eta_{\mu\nu}$ and $\eta^{\mu\nu}h_{\mu\nu}=0$. 
Expanding the graviton perturbations, we obtain
\beq
S^{(2)}&=&\int d^4 x\; a_0^2\left[-\frac14\parl_{\mu}h_{\r\sig}\parl^{\mu}
h^{\r\sig}+\frac12\parl_{\mu}h^{\r\nu}\parl_{\nu}h^{\mu}_{\ \r}
-\frac12\eta^{\mu\nu}\parl_{\mu}{\tilde \phi}\parl_{\nu}{\tilde
  \phi}\right.
\notag\\
&&\qquad\qquad +6\parl_{\mu}\omega\parl^{\mu}\omega
-\frac{24}{\tau^2}\omega^2-2\parl_{\mu}h^{\mu\nu}\parl_{\nu}\omega
-\frac{8}{\tau}\omega\parl_{\mu}h^{\mu0}+\frac{24}{\tau^2}h^{00}\omega\notag\\
&&\qquad \qquad +\frac{2}{\kappa\eta}\parl_{\nu}{\tilde
  \phi}\parl_{\mu}h^{\mu\nu}
+\frac{8}{\tau\kappa\eta}{\tilde \phi}\parl_{\mu}h^{\mu0}
-\frac{24}{\tau^2\kappa\eta}{\tilde \phi}h^{00}
-\frac{12}{\kappa\eta}\parl_{\mu}{\tilde \phi}\parl^{\mu}\omega\notag\\
&&\qquad\qquad \left.
-\left(\frac{48}{\tau^2\kappa\eta}+\frac{4\r\kappa}{H_0^2\tau^2}\right)
\omega {\tilde \phi} 
+\frac{1}{2}\left(\frac{48}{\tau^2\kappa^2\eta^2}
-\frac{\sig^2}{H_0^2\tau^2}\right){\tilde \phi}^2 
\right].
\eeq

There are the mixing terms among $h_{\mu\nu}$, $\omega$ and ${\tilde
  \phi}$. 
To obtain canonically normalized fields, let us
perform field redefinitions \cite{Ren:2014sya,Calmet:2013hia} as follows,
\beq
{\hat h}_{\mu\nu}=h_{\mu\nu}, \qquad {\omega}={\hat
  \omega}+ \frac{\zeta}{\kappa\eta}{\hat \phi},
\qquad {\tilde \phi}=\zeta{\hat \phi}, \label{canonical1}
\eeq
where $\zeta$ is given by
\beq
\zeta=\left(1+\frac{12}{\kappa^2\eta^2}\right)^{-1/2}
=\left(1+6\frac{M^2}{\eta^2}e^{-2v/\eta}\right)^{-1/2}
\eeq
and we have used the classical equation of motion 
\eqref{seq3}, which reads 
\beq \label{rkH}
\r+ \frac{24}{\kappa^2\eta}H_0^2=0.
\eeq
when expressed in terms of $\r$, $\kappa$ and $H_0$.
In terms of the new fields, we have
\beq
S^{(2)}\hspace{-0.2cm}&=&\hspace{-0.2cm}\int d^4 x 
a_0^2\left[-\frac14\parl_{\mu}h_{\r\sig}\parl^{\mu}h^{\r\sig}+
\frac12\parl_{\mu}h^{\r\nu}\parl_{\nu}h^{\mu}_{\ \r}
+6\parl_{\mu}{\hat \omega}\parl^{\mu}{\hat \omega}
-\frac{24}{\tau^2}{\hat \omega}^2 
\right.\notag\\
&&\qquad \left.
-2\parl_{\mu}h^{\mu\nu}\parl_{\nu}{\hat\omega}
-\frac{8}{\tau}{\hat\omega}\parl_{\mu}h^{\mu0}
+\frac{24}{\tau^2}h^{00}{\hat\omega}
-\frac12\eta^{\mu\nu}\parl_{\mu}{\hat \phi}\parl_{\nu}{\hat \phi}
-\frac12a_0^2m^2_{\hat \phi}{\hat \phi}^2 
\right],
\eeq
where the mass of ${\hat \phi}$ is
\beq
m^2_{\hat \phi}=\zeta^2\left(\sig^2+\frac{4\r}{\eta}\right).
\eeq

The quadratic action $S^{(2)}$ is now written in 
the standard 
canonical form for graviton perturbations and a massive scalar field 
with mass $m_{\hat \phi}$. It is known that the propagator for
massive 
scalar field in de Sitter space is given by the hypergeometric function 
\cite{Chernikov:1968zm}
\beq
\langle {\hat \phi}(x){\hat \phi}(x') \rangle=\frac{H_0^2}{(4\pi)^2}
\Gamma\left(\frac{3}{2}+\nu\right)\Gamma\left(\frac{3}{2}-\nu\right)
{}_2F_1\left(\frac{3}{2}+\nu,\frac{3}{2}-\nu;2;1-\frac{y}{4}\right)
\label{smpropagator}
\eeq
with $\nu=\sqrt{\frac{9}{4}-\frac{m_{\hat \phi}^2}{H_0^2}}$,
where the de Sitter invariant length $y$ is defined by
\beq
y\equiv \frac{-(\tau-\tau')^2+({\vec x}-{\vec x}')^2}{\tau\tau'}.
\eeq
For the graviton propagator, let us introduce the following form of
the  gauge fixing term \cite{TW94}
\beq
{\cal L}_{\rm GF}=-\frac12 a_0^2 \eta^{\mu\nu}F_{\mu}F_{\nu},\quad
F_{\mu}
=\parl_{\r}h_{\mu}^{\, \r}-2\parl_{\mu}{\hat\omega} 
-\frac{2}{\tau}h_{\mu}^{\, 0}-\frac{4}{\tau}\delta^{0}_{\mu}{\hat\omega}. 
\eeq
The quadratic terms of the graviton perturbations in the gauge-fixed action 
are
\beq
S^{(2)}_{\rm gr}&=&\int d^4 x  \; a_0^2\left[\ 
\frac{1}{2}\eta^{\mu\nu}\partial_\mu X\partial_\nu X
-\frac{1}{4}\eta^{\mu\nu}\partial_\mu{\bar h}^i_{\ j}
\partial_\nu{\bar h}^j_{\ i}
+\frac{1}{2}\eta^{\mu\nu}\partial_\mu h^{0i}\partial_\nu h^{0i}
\right.\notag\\
&&\qquad \qquad \qquad \left.
+a_0^2H_0^2h^{0i}h^{0i}
-\frac{1}{2}\eta^{\mu\nu}\partial_\mu Y\partial_\nu Y 
-a_0^2H_0^2Y^2\right],\label{quad}
\eeq
where ${\bar h}^i_{\ j}$, $X$ and $Y$ are defined by 
\beq
&&{\bar h}^i_{\ j}\equiv h^i_{\ j}-\frac13h^k_{\ k}\delta^i_{\ j}
=h^i_{\ j}-\frac13h^{00}\delta^i_{\ j},\quad {\bar h}^i_{\ i}=0,\\
&&X\equiv 2\sqrt{3}{\hat\omega}-\frac{1}{\sqrt{3}}h^{00},
\hspace{1em}Y\equiv h^{00}-2{\hat\omega}. 
\eeq
Note that 
the fields ${\bar h}^i_{\ j}$ and $X$ satisfy the same equation of
motion 
as the massless minimally coupled scalar field; while the fields  
$h^{0i}$ and $Y$ satisfy the same equation of motion as the massless 
conformally coupled scalar field. 
As a result, their propagators are given by
\beq
\begin{split}
\langle{\bar h}^i_{\ j}(x){\bar h}^k_{\ l}(x')\rangle&
=\left(\delta^{ik}\delta_{jl}+\delta^i_{\ l}\delta_j^{\ k}
-\frac{2}{3}\delta^i_{\ j}\delta^k_{\ l}\right)
\langle\varphi(x)\varphi(x')\rangle,\\ 
\langle X(x)X(x')\rangle&=-\langle\varphi(x)\varphi(x')\rangle, 
\end{split}
\eeq
where $\varphi$ denotes a massless minimally coupled scalar field, 
\be
\langle\varphi(x)\varphi(x')\rangle
=\frac{H_0^2}{4\pi^2}\left(\frac{1}{y}-\frac{1}{2}\ln y
+\frac{1}{2}\ln (a_0 (\tau)a_0(\tau'))+1-\gamma\right); \label{mmc}
\ee
and 
\beq
\begin{split}
\langle h^{0i}(x)h^{0j}(x')\rangle&=-\delta^{ij}
\frac{H_0^2}{4\pi^2}\frac{1}{y}, \quad \langle Y(x)Y(x')\rangle
=\frac{H_0^2}{4\pi^2}\frac{1}{y}. \label{pro1}
\end{split}
\eeq
In appendix B, we make comments on the IR properties of the two-point
functions of massless minimally coupled scalar fields in de Sitter and
flat FRW universe.

We will be computing the quantum shift of the vev of the scalar field
by looking at its 1-loop tadpole with various modes running in the
loop. It is well known that the de Sitter breaking 
IR  logarithm are generated by  massless minimally coupled scalar
field, but not by  conformally coupled fields, therefore we can neglect the 
propagation of the  (massive) scalar field $\phi$ and the  
conformally coupled fields $Y$ and $h^{0i}$; and concentrate on the
propagation of the modes ${\bar h}_{ij}$ and $X$. In this approximation 
\cite{Kitamoto:2012ep,Kitamoto:2014gva}, 
\beq
h^{00}\simeq2{\hat\omega}\simeq \frac{\sqrt{3}}{2}X.\label{ncon}
\eeq  
and we get the following three propagators.
\beq
\begin{split}
\langle{ h}^i_{\ j}(x){
  h}^k_{\ l}(x')\rangle&\simeq(\delta^{ik}\delta_{jl}
+\delta^i_{\ l}\delta_j^{\ k}
-\frac{3}{4}\delta^i_{\ j}\delta^k_{\ l})\langle\varphi(x)\varphi(x')\rangle,\\ 
\langle h^{00}(x){ h}^i_{\ j}(x')\rangle&\simeq
-\frac14 \delta^i_{\ j}\langle\varphi(x)\varphi(x')\rangle,\\
\langle h^{00}(x){ h}^{00}(x')\rangle&\simeq
-\frac34 \langle\varphi(x)\varphi(x')\rangle.\label{pro3}
\end{split}
\eeq
As we are interested in the de Sitter breaking IR logarithm, let us
retain only the  IR logarithmic parts in \eq{mmc}.
We obtain in this approximation, 
\beq
\begin{split}
\langle{ h}^i_{\ j}(x){ h}^k_{\ l}(x')\rangle&\simeq
(\delta^{ik}\delta_{jl}+\delta^i_{\ l}\delta_j^{\ k}
-\frac{3}{4}\delta^i_{\ j}\delta^k_{\ l}) \frac{H_0^2}{8\pi^2}
\ln (a_0(\tau)a_0(\tau')),\\ 
\langle h^{00}(x){ h}^i_{\ j}(x')\rangle&\simeq
- \delta^i_{\ j} \frac{H_0^2}{32\pi^2}\ln (a_0(\tau)a_0(\tau')),\\
\langle h^{00}(x){ h}^{00}(x')\rangle&\simeq
- \frac{3H_0^2}{32\pi^2}\ln (a_0(\tau)a_0(\tau')).
\end{split}
\eeq
Below we will show that this IR logarithm of the graviton loop
in de Sitter space could have a screening effect on the cosmological constant.

\subsection{Graviton loop corrections to $v$}

\begin{figure}[t]
 \centering
 \vspace{-0cm}
  \includegraphics[width=10cm]{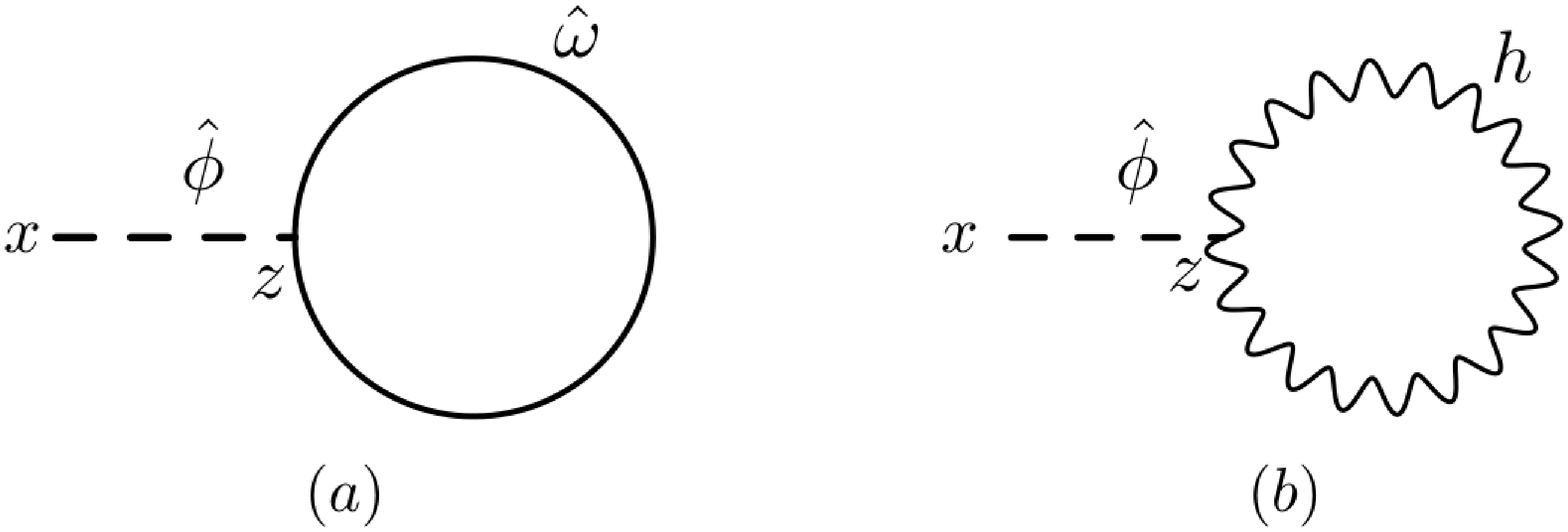}
  \vspace*{-0cm} 
\caption{ The one-loop tadpole diagrams due to the graviton loops. 
The broken line denotes the scalar ${\hat \phi}$ propagator, 
the solid line in $(a)$ denotes the ${\hat \omega}$ propagator and the
wavy line in $(b)$ denotes the $h$ propagators.}
 \vspace{0cm}
 \label{fig:one}
\end{figure}

Now we evaluate the graviton one-loop corrections to the vev of the
dilaton scalar field in the de Sitter background \eq{seq3}, \eq{dscur}.  
The diagrams to be considered here are shown in
Fig. \ref{fig:one}. We have expanded the scalar field around the
zeroth 
order vev,
$v$, then the one-loop tadpole diagrams generating the vev of the
shifted scalar field ${\tilde \phi}$ correspond to a (time-dependent)
shift of $v$ \cite{Weinberg:1973ua}. From \eqref{canonical1}, the vev
of $\tilde{\phi}$ is related to that of ${\hat \phi}$ as
\beq
 \langle \Omega| {\tilde \phi}(x)|\Omega \rangle
=\zeta\langle\Omega| {\hat \phi}(x)|\Omega\rangle. \label{relation1}
\eeq
The interaction vertices that are relevant to the tadpole diagrams are
types of ${\hat \phi}hh$ and ${\hat \phi}{\hat \omega}{\hat \omega}$.
The corresponding interaction Lagrangian is obtained from
\eqref{int12} and \eqref{int32} given in appendix A:  
\beq \label{tadint}
\frac{1}{a_0^2{\hat \phi}}{\cal L}_{\rm tad} =  -\frac{2\zeta}{\eta}
\left(-\frac14\parl_{\mu}h_{\r\sig}\parl^{\mu}h^{\r\sig}
-\frac12\parl_{\mu}(h^{\mu}_{\ \r}\parl_{\nu}h^{\r\nu})
-\frac12h^{\mu}_{\ \r}\parl_{\mu}\parl_{\nu}h^{\r\nu}
+\frac{3}{\tau}\parl_{\mu}(h^{0\nu}h_{\nu}^{\ \mu})
\right.\notag\\ \qquad\qquad \left.
-\frac{6}{\tau^2}h^{\r0}h_{\r}^{\ 0} -6\parl_{\mu}({\hat
  \omega}\parl^{\mu}{\hat \omega}) -6{\hat
  \omega}\parl_{\mu}\parl^{\mu}{\hat \omega} - \frac{24}{\tau}{\hat
  \omega}\parl_0{\hat \omega} +\frac{24}{\tau^2}{\hat \omega}^2\right)
-8a_0^2\r\kappa^2\zeta {\hat \omega}^2 . \;\; 
\eeq

We will use the
in-in formalism (Schwinger-Keldysh formalism)
\cite{Chou:1984es,Calzetta:1986ey,Jordan:1986ug,Weinberg:2005vy} to
calculate the vev of ${\hat \phi}$ at the one-loop level. In the in-in
formalism only "in" vacuum is prepared (as the Bunch-Davies vacuum in
our case) and the vev of 
operators with respect to the in state can
be evaluated. The two time sheets which we will call $+$ and $-$
sheets are introduced. We shall put $x$ at which the operator ${\hat
  \phi}$ is inserted on the $+$ sheet. We should take into account
the possibility that the vertices at $z$ can be of $+$ type or $-$
type. The in state is developed along the $+$ sheet by the time
evolution operator of $+$ type and then goes back in time along the
$-$ sheet by the time evolution operator of $-$ type. Making use of
the in-in formalism the vev of ${\hat \phi}_+(x)$ is given by  
\beq
\langle\Omega| {\hat \phi}_+(x)|\Omega\rangle= {\langle 0| {\tilde
    T}\{e^{i\int_{\tau_i}^0 H_{{\rm int}-}d\tau'}\} {T}\{{\hat
    \phi}_{+}(x)e^{-i\int_{\tau_i}^0 H_{{\rm int}+}d\tau''}\} |0
  \rangle}, 
 \label{vevinin}
\eeq
where the amplitudes of vacuum bubbles always cancel in the in-in formalism. 
We have introduced the initial time $\tau_i$. 
It can be seen from (\ref{vevinin}) that contributions from
$\tau'>\tau$ 
automatically cancel and 
\eqref{vevinin} is equivalent to 
\be
\langle\Omega| {\hat \phi}_+(x)|\Omega\rangle=\langle 0| 
{\tilde T}\{e^{i\int_{\tau_i}^{\tau} H_{{\rm int}-}d\tau'}\}
{\hat \phi}_{+}(x){T}\{e^{-i\int_{\tau_i}^{\tau} H_{{\rm
      int}+}d\tau''}\} |0 \rangle
.\notag
\ee
The distance $y$ is determined by the two space-time points 
and becomes $y_{ab}\equiv y(x_a,x'_b)$ $(a,b=+,-)$:
\beq
\begin{split}
y_{++}(x,x') &\equiv H_0^2a_0(\tau)a_0(\tau ')\Delta x_{++}^2 =
H_0^2a_0(\tau)a_0(\tau ')  
[({\vec x}-{\vec x}')^2 -(|\tau-\tau'|-i\delta)^2],\\
y_{+-}(x,x') &\equiv H_0^2a_0(\tau)a_0(\tau ')\Delta x_{+-}^2 =
H_0^2a_0(\tau)a_0(\tau ') 
 [({\vec x}-{\vec x}')^2 -(\tau-\tau'+i\delta)^2],\\
y_{-+}(x,x') &\equiv H_0^2a_0(\tau)a_0(\tau ')\Delta x_{-+}^2 =
H_0^2a_0(\tau)a_0(\tau ')  
[({\vec x}-{\vec x}')^2 -(\tau-\tau'-i\delta)^2],\\
y_{--}(x,x') &\equiv H_0^2a_0(\tau)a_0(\tau ')\Delta x_{--}^2 =
H_0^2a_0(\tau)a_0(\tau ')  
[({\vec x}-{\vec x}')^2 -(|\tau-\tau'|+i\delta)^2],
\end{split}
\eeq
where $\Delta x^2=\eta_{\mu\nu}\Delta x^{\mu}\Delta x^{\nu}$ with 
$\Delta x^{\mu}=(x-x')^{\mu}$ and $\delta >0$ is infinitesimal.
Depending on the kinds of $y$, we have four kinds of propagators 
for ${\hat \phi}$:
\beq
\begin{split}
i\Delta_{++}(x,x')&\equiv 
\langle T\{{\hat \phi}(x_+){\hat \phi}(x'_+)\} \rangle ,\\
 i\Delta_{+-}(x,x')&\equiv
 \langle {\hat \phi}(x'_-){\hat \phi}(x_+) \rangle ,\\
 i\Delta_{-+}(x,x')&\equiv 
 \langle {\hat \phi}(x_-){\hat \phi}(x'_+) \rangle,\\
 i\Delta_{--}(x,x')&\equiv
 \langle {\tilde T}\{{\hat \phi}(x_-){\hat \phi}(x'_-)\} \rangle.
\end{split}
\eeq
Note that the operators on $-$ sheet are always in the left side of
that on $+$ sheet in correlation functions. In 
our case of the one-loop tadpole, 
we need the first two types of the scalar field propagators.
For the graviton propagator we only need its short distance limit in
evaluation of the one-loop tadpole diagrams. The coincidence limit
gives the same result regardless of its type ($\pm$).

In the in-in formalism, the first order correction to (\ref{vevinin}) is 
\beq
&& \hspace{-0.5cm}\langle\Omega| {\hat \phi}_+(x)|\Omega\rangle\notag\\
\hspace{-0.2cm}&=&\hspace{-0.2cm}
{\langle 0| {\tilde T}\{1+{i\int_{\tau_i}^0 H_{{\rm int}-}d\tau'}\}
{T}\{{\hat \phi}_{+}(x)(1{-i\int_{\tau_i}^0 H_{{\rm int}+}d\tau''})\} 
|0 \rangle}\notag\\
\hspace{-0.2cm}&=&\hspace{-0.2cm}
-i\int_{\tau_i}^0 d\tau'\int d^3 z'
{\langle 0| {\tilde T}\{{ {\cal L}_{{\rm int}-}(z')}\}
{\hat \phi}_{+}(x)|0 \rangle}
+i\int_{\tau_i}^0 d\tau'' \int d^3 z''
{\langle 0|{T}\{ {\hat \phi}_{+}(x)
{ {\cal L}_{{\rm int}+}(z'')}\} |0 \rangle}
\notag\\
\label{tadfirst}
\eeq
Substituting (\ref{tadint}) into (\ref{tadfirst}) yields
\be
\langle\Omega| {\hat \phi}_+(x)  |\Omega\rangle   
= i\frac{2\zeta}{\eta}\int_{\tau_i}^0 d\tau'\int d^3 z'
 a_0(\tau')^2 (i\Delta_{+-}(x,z')-i\Delta_{++}(x,z')) \times I,
\ee
where 
\beq
 I 
:= \hspace{-0.5cm} && 
-\frac14 \lim_{w \to z' }\parl'_{\mu}\parl^{\mu}
\langle h_{\r\sig}(z')h^{\r\sig}(w)\rangle
-\frac12\parl'_{\mu}(\lim_{w \to z' }\parl_{\nu} \langle 
h^{\mu}_{\ \r}(z')h^{\r\nu}(w)\rangle) 
\notag\\
&&
-\frac12\lim_{w \to z' }\parl_{\mu}\parl_{\nu}
\langle h^{\mu}_{\ \r}(z')h^{\r\nu}(w)\rangle
+
\frac{3}{\tau'}\parl'_{\mu}\langle h^{0\nu}(z')h_{\nu}^{\ \mu}(z')\rangle
-\frac{6}{\tau'^2}\langle h^{\r0}(z')h_{\r}^{\ 0}(z')\rangle
\notag\\
&& 
-6\parl'_{\mu}(\lim_{w \to z' }\parl^{\mu}
\langle {\hat \omega}(z'){\hat \omega}(w)\rangle)
-6\lim_{w \to z' }\parl_{\mu}\parl^{\mu}
\langle{\hat \omega}(z'){\hat \omega}(w)\rangle
- \frac{24}{\tau'}\lim_{w \to z' }\parl_0
\langle{\hat \omega}(z'){\hat \omega}(w)\rangle
\notag\\
&&
+\frac{24}{\tau'^2}
\langle{\hat \omega}(z'){\hat \omega}(z')\rangle
+4a_0(\tau')^2\r\kappa^2\eta  \langle{\hat \omega}(z'){\hat
  \omega}(z')\rangle.
    \label{tadfirst2}
\eeq 
Here $\parl'$ acts on $z'$ and $\parl$ acts on $w$. 
To obtain the time dependent effects, 
the graviton propagators are approximated by the IR 
logarithms and then only the zeroth component of the derivatives 
on the propagators can contribute. We neglect $h^{0i}$ and $Y$
components 
of the graviton perturbation and make use of (\ref{ncon}) as mentioned
previously. 
Then (\ref{tadfirst2}) is evaluated as 
\beq
&& \hspace{-1cm}\langle\Omega| {\hat \phi}_+(x)|\Omega\rangle\notag\\
 &=&- i\zeta \frac{3H_0^4}{4\pi^2\eta}\int_{\tau_i}^0 d\tau'\int d^3 z'
 a_0(\tau')^4 (i\Delta_{+-}(x,z')-i\Delta_{++}(x,z')) \left(1-6\ln
 a_0(\tau')\right), \qquad
 \label{tadfirst4}
\eeq 
where we have used the classical equation of motion \eq{rkH}.
Now it is clear that the dimensionless coupling constant for the
interaction of the graviton and the dilaton scalar field is given by
$H_0\zeta/\eta$.

Next let us consider the part $i(\Delta_{+-}-\Delta_{++})$  in
(\ref{tadfirst4}). The propagator is given in (\ref{smpropagator}) and
can be rewritten as \cite{Bunch:1978yq}
\beq
&&\hspace{-1cm}i\Delta(x,z')\notag\\
&=&\hspace{-0.3cm}\frac{H_0^2}{(4\pi)^2}
\left[\frac{4}{y}
+\frac{1}{\Gamma \left(\frac{1}{2}+\nu\right)\Gamma \left(\frac{1}{2}-\nu\right)}
\sum _{n=0}^{\infty}
\frac{\Gamma \left(\frac{3}{2}+\nu+n\right) \Gamma
   \left(\frac{3}{2}-\nu+n\right)}{n! (n+1)!}
   \left(\frac{y}{4}\right)^n
    \right.\notag\\
   &&\left.\hspace{-0.3cm} \times\left(\psi \left(\frac{3}{2}+\nu+n\right)
   +\psi \left(\frac{3}{2}-\nu+n\right)-\psi (n+1)-\psi
   (n+2)+\ln \left(\frac{y}{4}\right)\right)\right]. \qquad 
\label{smpropagator2}
\eeq
It includes a infinite sum and is not easy to handle exactly. 
To avoid this difficulty, we restrict ourselves to small mass case, 
$m^2_{\hat \phi}\ll H_0^2$, where the expression (\ref{smpropagator2}) 
can be reduced to
\beq
i\Delta(x,z')=\frac{3 H_0^4}{8 \pi ^2 m_{\hat \phi}^2}+\frac{H_0^2}{4 \pi
  ^2 y} -\frac{H_0^2}{8 \pi ^2} \ln
   \left(\frac{y}{4}\right)-\frac{7 H_0^2}{24 \pi ^2}+{\cal O}(m_{\hat
     \phi}^2). 
\label{smpropagator3}
\eeq
Then $i(\Delta_{+-}-\Delta_{++})$ can be simplified to
\beq
i\Delta_{+-}(x,z')-i\Delta_{++}(x,z')
=\frac{H_0^2}{4\pi^2}\left(\frac{1}{y_{+-}}-\frac{1}{y_{++}}\right)
-\frac{H_0^2}{8\pi^2}(\ln y_{+-}-\ln y_{++}).\label{subtraction}
\eeq
In appendix \ref{app-m2}, we explore the parameter region where 
$m^2_{\hat \phi}\ll H_0^2$ is satisfied.

To evaluate the integral in \eq{tadfirst4}, we follow the trick of 
\cite{Onemli:2002hr} and  use the identity 
\beq
  \parl^2 \ln(\Delta x^2)
=(-\parl_0^2+\parl_i^2)\ln(\Delta x^2)=\frac{4}{\Delta
  x^2}, \label{identity}
\eeq 
the integral of the first term can then be written as
\be
 \int d^4z'a_0(\tau')^4 \left(\frac{1}{y_{+-}}-\frac{1}{y_{++}}\right)
= \frac{1}{H_0^2a_0(\tau)}\parl^2 \int d^4z' a_0(\tau')^3 
\left(\ln(\Delta x_{+-}^2)-\ln(\Delta x_{++}^2)\right), \quad 
\ee
where we have used
\beq
\int^{0}_{\tau_i}d\tau'\int^{\infty}_{-\infty} d^3z'\parl^2
=\parl^2\int^{0}_{\tau_i}d\tau'\int^{\infty}_{-\infty} d^3z'.
\eeq
Now we are left with the logarithms
\beq
\ln \Delta x_{++}^2&=&\ln (r^2-\Delta \tau^2+2i\delta |\Delta \tau|),\\
\ln \Delta x_{+-}^2&=&\ln (r^2-\Delta \tau^2-2i\delta \Delta \tau),
\eeq
with $\Delta \tau=\tau-\tau'$. $\ln x$ has a cut for $x<0$, then 
\beq
\ln \Delta x_{++}^2&=&\ln |r^2-\Delta \tau^2|+i\pi \theta(\Delta\tau^2-r^2),\\
\ln \Delta x_{+-}^2&=&\ln |r^2-\Delta \tau^2|-i\pi
\theta(\Delta\tau^2-r^2)
(\theta(\Delta\tau)-\theta(-\Delta\tau)),
\eeq
gives
\beq
\ln \Delta x_{+-}^2-\ln \Delta
x_{++}^2&=&(-i\pi\theta(\Delta\tau^2-r^2))
(1+\theta(\Delta\tau)-\theta(-\Delta\tau)),\label{sub2}
\eeq
where $\theta(x)$ is $1$ for $x>0$ and $0$ for $x<0$. 
As is seen from (\ref{sub2}), we only have the contribution from 
inside of the past light cone where $r^2\leq\Delta\tau^2$ and
$\tau\geq\tau'$ are satisfied.

The vertex integrations can then be performed straightforwardly. 
As a result, we have 
\beq
&& \hspace{-1cm}\langle\Omega| {\hat \phi}_+(x)|\Omega\rangle\notag\\
&=&\hspace{-0.2cm} 
-i \frac{3H_0^6\zeta}{16\pi^4\eta} \int d^4 z'
 a_0(\tau')^4 \left(\frac{1}{y_{+-}}-\frac{1}{y_{++}}
-\frac12\left(\ln y_{+-}-\ln y_{++}\right)\right)
  \left(1-6\ln a_0(\tau')\right)\notag\\
&=&\hspace{-0.2cm} 
-\frac{H_0^2\zeta}{2\pi^2\eta}
\left(\frac{3}{2}\ln^2 a_0(\tau)+13\ln a_0(\tau)
-16+\frac{39}{2a_0(\tau)}
-\frac{37}{12a_0(\tau)^2}-\frac{1}{6a_0(\tau)^3}
\right),
\label{tadpoleresult}
\eeq
where we have chosen, for convenience, 
the initial time as $\tau_i=-1/H_0$ ($t_i=0$). 
It is equivalent to taking the initial size 
of the universe as $L_0=-H_0^{-1}$.
We focus on the leading contribution to the vev with respect to the IR logarithm,
\beq
\frac{\d v}{\zeta} := \langle\Omega| {\hat \phi}_+(x)|\Omega\rangle
&\sim& -\frac{\zeta}{\eta}\frac{3H_0^2}{4\pi^2}\ln^2 a_0(\tau).\label{tadleading}
\eeq
From (\ref{relation1}), we finally obtain
\bea \label{vtot-1}
v_{\rm eff}(\tau)
&=&v+ \d v \nn\\
&\sim&v- \frac{\zeta^2}{\eta}\frac{3H_0^2}{4\pi^2}\ln^2 a_0(\tau).\label{effvev}
\eea

It is instructive to rewrite the above result in the form
\be
v_{\rm eff}(\tau)=v(1 - \epsilon \, \ln^2 a_0(\tau)
+\mbox{sub-leading}),\label{veffdrg}
\ee
with
\be
\epsilon\equiv \frac{\zeta^2}{v\eta}\frac{3H_0^2}{4\pi^2}. 
\ee
It is clear that the 1-loop result \eq{veffdrg} is valid only for
$|\e \, \ln^2 a_0|\ll 1$ .
In particular, the result \eq{veffdrg} 
cannot be applied when  $a_0(\t)$ has grown significantly
after a sufficiently long time.  
The presence of these large IR effects indicates that the
perturbative calculation eventually breaks down and a resummation of
the secular terms  is called for.  As we discussed in the
introduction, we currently do not have a good
enough understanding of quantum gravity and the technology to
determine precisely the behaviour of the secular IR effects of
graviton loops.  
In this paper we will employ the DRG method \cite{burgess,bur1}
 to improve the result \eqref{effvev}. 
The DRG effectively resums the leading order time-dependent
corrections and requires only small $\e$ in order to be valid. This 
is analogous 
to the renormalization group improvement of the perturbative correction from leading
logarithm of $\ln \mu$. 
We believe the time
dependent behaviour we found are qualitatively correct although the
precise details may be different. However this must be checked. 
In any case, as we will see in section 3.3, the screening of the cosmological constant
 weakly depends on the classical vev $v$, only through $\zeta$.
  Thus in principle we can have a large screening from our one-loop result
 \eqref{veffdrg} by choosing $v$ large enough such that $|\epsilon \ln^2 a_0|\ll 1$
  without relying on the resummation method.

The result of  DRG is
\be \label{dRG-v}
v_{\rm eff} (\t) = r + \frac{1}{q} W(pq e^{-qr})
\ee
where
\be \label{abc}
p:= \frac{3M^2}{\eta}, 
\qquad q : = \frac{2}{\eta},\qquad  
r:= v- \frac{3M^2}{\eta} e^{-2 v/\eta} - \frac{3 H_0^2 \ln^2 a_0(\t)
}{4 \pi^2 \eta} 
\ee
and $W(z)$ is the Lambert-$W$ function, which is defined to be the
analytic function  that satisfies the relation 
\be
z = W(z) e^{W(z)}.
\ee
Here the principal branch is used.
The result \eq{dRG-v} is valid as long as $\e \ll 1$. 
As a consistency check, let us write $r$ as
\be
r =  \underbrace{v- \frac{3M^2}{\eta} e^{-2 v/\eta}}_{r_0} 
- \underbrace{\frac{3 H_0^2 \ln^2 a_0(\t)}{4 \pi^2 \eta}}_{r_1}
:= r_0- r_1
\ee
and note that for small $\updelta$, the Lambert-$W$
function has the expansion
\be \label{W-exp}
W(z(1+ \updelta)) = \a + \frac{\a}{1+\a} \updelta  + O(\updelta^2), \quad 
\mbox{where} \quad \a := W(z).
\ee
It is then straightforward to verify that \eq{dRG-v}
reproduces the 1-loop result \eq{veffdrg} in the
limit of small $qr_1$, noting $W(z e^z)=z$.

Our result
reveals that a time-dependent vev of
the scalar field $\phi$ is generated through de Sitter symmetry 
breaking IR logarithm contained in the graviton one-loop corrections
\be
v_{\rm eff}(\t) = v + \d v (\t).
\ee 
At the tree level, the Hubble parameter is given by
\be \label{H0a0}
H_0^2 = \frac{1}{3M^2} e^{\frac{2v}{\eta}} V(v) \quad \mbox{and}
\quad
a_0(t) = e^{H_0 t}. 
\ee
The 1-loop IR effects of the graviton loops induces a time dependent
shift $\d v$ in the vev and modifies the Hubble parameter in the string
frame to
\bea \label{H2}
H^2 &=&  \frac{1}{3M^2} e^{\frac{2v_{\rm eff}}{\eta}} V(v) \nn\\
&= &
H_0^2 e^{\frac{2\d v}{\eta}}
\eea
which is now time dependent. 
Note that in the first equality we have taken into account the time 
dependence only in the vev in the exponent, and we have 
neglected that in the dilaton potential. Actually,  as we will discuss
in section 3.1 for the Einstein frame  Hubble parameter , we do not need to 
know the precise form of the time dependent IR effects on the 
dilaton potential in order to have \eq{H2}. 

\begin{figure}[h!] 
\centering
\includegraphics[width=10.0cm]{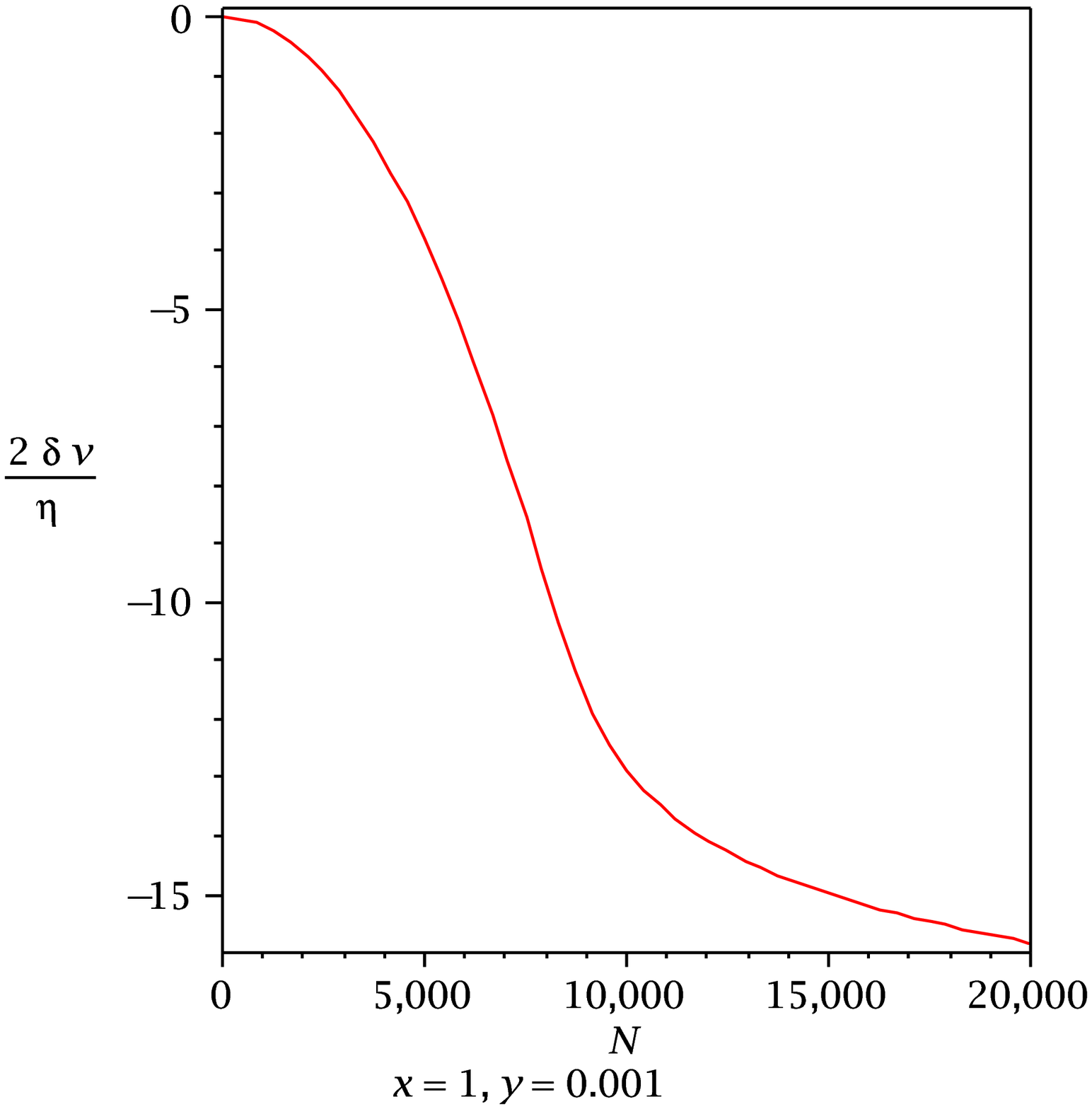}
\caption{Plot of  $\frac{2\delta v}{\eta}$ against $N$}
\label{fig:two}
\end{figure}

It is interesting to see how the time dependence looks like. Let us
introduce the quantity
\be
\cR := e^{\frac{2  \d v}{\eta}},
\ee
which will turn out to be useful later.
Using \eq{dRG-v}, it is easy to verify that
\be \label{RW}
\cR = \frac{6 y^2}{W(z)},
\ee
where
\be \label{z}
z: = 6 y^2 e^{6y^2 (1+ \frac{N^2}{4\pi^2 x^2})}
\ee
and
\be \label{xy}
x:=\frac{M e^{-v/\eta}}{H_0}, \quad y:=\frac{M e^{-v/\eta}}{\eta}.
\ee
Here  $N$ is defined by
\be \label{N-string}
N :=\ln a_0 = H_0 t
\ee 
and is simply the time measured in unit of $H_0^{-1}$.
For a given model, the parameters $x$ and $y$ are fixed, and time
dependence enters through the
parameter $z$. 
At initial times where $N$ is small, it is
\be
W(z) = 6 y^2 + \frac{36 y^4}{1+6 y^2} \frac{N^2}{4 \pi^2 x^2} + \cO(N^4)
\ee
and
\be
\cR =1 - \frac{6y^2}{1+6 y^2} \frac{N^2}{4 \pi^2 x^2}+ \cO(N^4).
\ee
In this regime, $2 \d v/\eta \propto - N^2$ decreases with $N$ quadratically.  
A later times when $N$ becomes large, it is
\be
W(z) = \frac{3 y^2}{2\pi^2 x^2} N^2 +\cO(\ln N),
\ee
and 
\be
\cR = \frac{4 \pi^2 x^2}{N^2} (1 + \cO( \frac{\ln N}{N}))
\ee
This means
\be
\frac{2 \delta v}{\eta} \simeq - \ln (\frac{N^2 H_0^2}{4\pi^2 M^2 e^{-2v/\eta}}) 
\ee
for late time. 
A typical plot of $2\delta v/\eta$ as a function of $N$ is shown in 
Fig.\ref{fig:two}. 

\subsection{Backreaction}

Our  result \eq{H2} about the 1-loop corrected Hubble parameter 
is based on perturbative quantum field theory 
in a (fixed) de Sitter background. 
Since  $H$  changes with time,  our computation is valid only if the
background is changing not too rapidly so that one can trust the
quantum field theory computation. 
One may introduce 
the ``slow roll parameter''
\be
\uve:= \frac{d }{dt} H^{-1}
\ee
as in the inflationary scenario. 
Here $t$ is the time in the tree level string frame metric
\eq{g0-string}. Note that $\uve$ may also be written as
\be
\uve = \frac{d \ln H }{d\cN},
\ee
where $d \cN := H dt$ measures the number of e-folding. 
In this representation, $\uve$ measures the fractional change of the Hubble
parameter per  e-folding. Obviously we
want $\uve \ll 1$ in order for our perturbative computation to be
trustable. However  this is not the only
requirement. We also want the accelerative change of the
metric to be small since this effect would backreact directly on the
solution through the Einstein equation. Following standard inflation, 
let us introduce the parameter
\be
\upeta : = \frac{ d \ln \uve}{d \cN} ,
\ee
which measure the fractional change of $\uve$ per e-folding. In general
we need 
\be \label{limit-s}
\uve, \upeta \ll 1
\ee
in order to trust the quantum field theory computation.

\begin{figure}[ht!] 
\centering
\begin{minipage}[b]{0.47\linewidth}
\includegraphics[width=8.0cm]{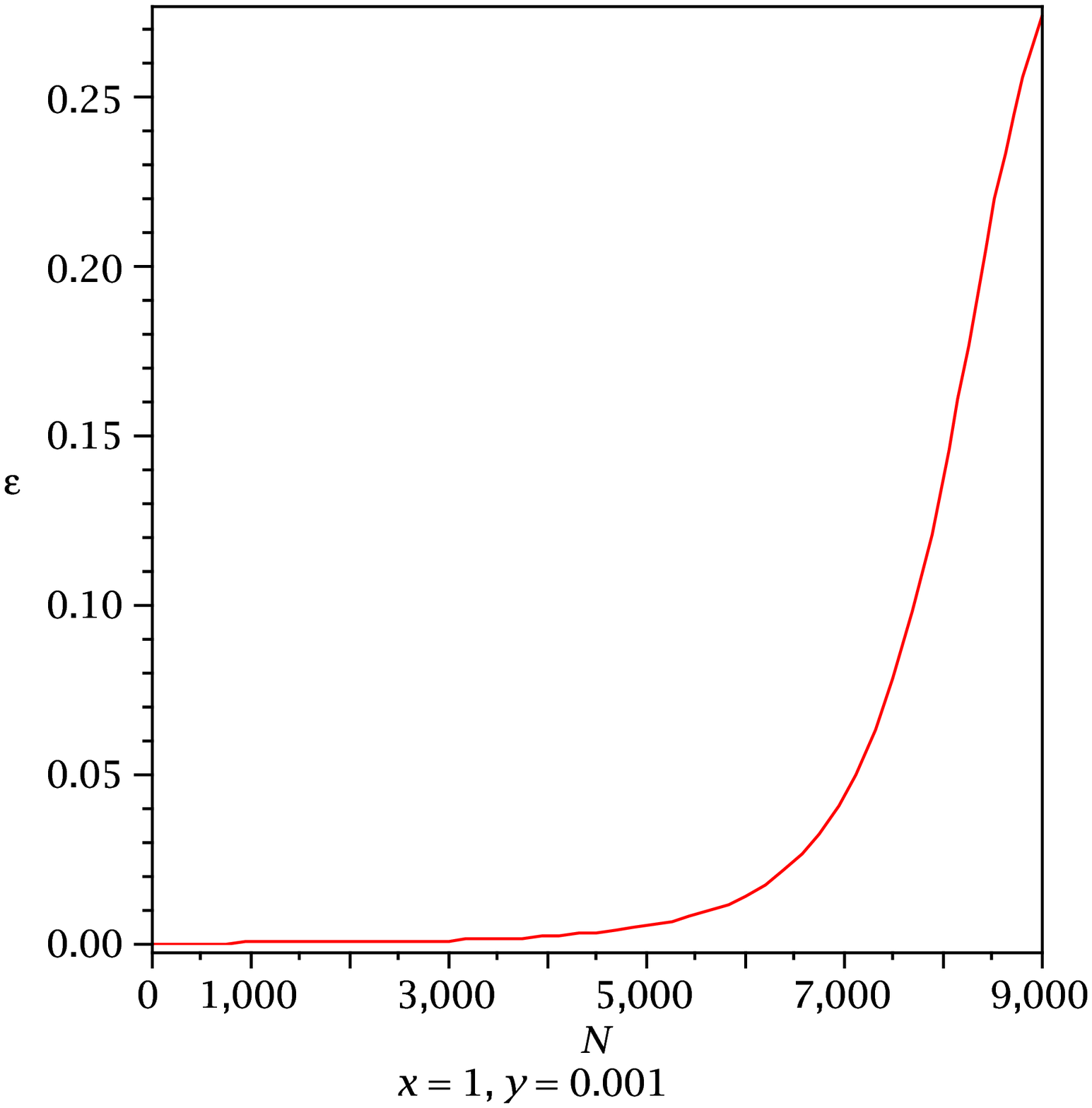}
\caption{Plot of  $\uve$ against $N$}
\label{e-s}
\end{minipage}
\quad
\begin{minipage}[b]{0.47\linewidth}
\includegraphics[width=8.0cm]{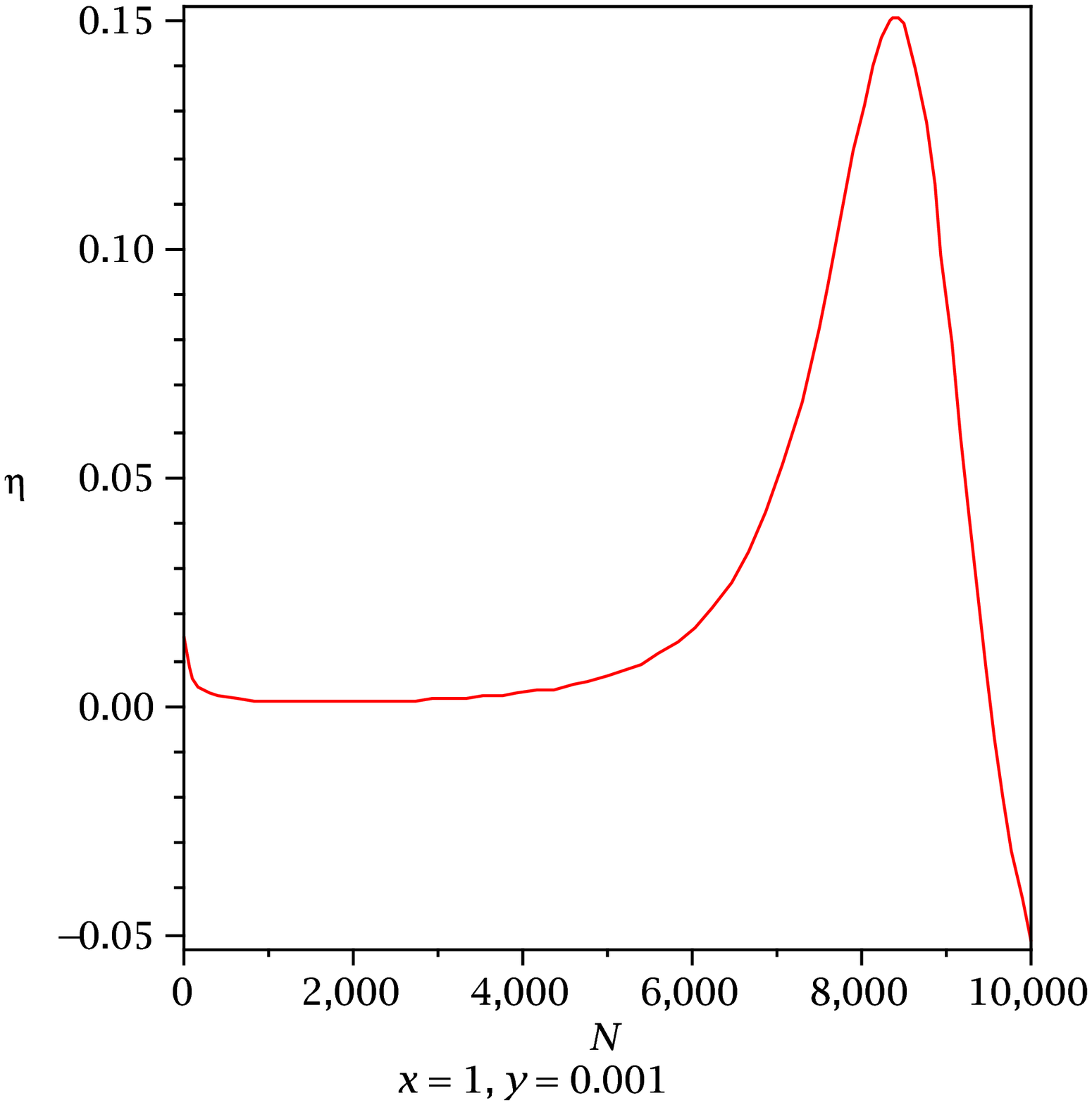}
\caption{Plot of  $\upeta$ against $N$}
\label{eta-s}
\end{minipage}
\end{figure}

It is easy to compute $\uve$ and $\upeta$ for our model. 
Substituting \eq{dRG-v} and
\eq{H2}, we obtain
\be \label{slow_roll-string}
\uve = \frac{3}{2\sqrt{6}\pi^2} \cdot \frac{N y \sqrt{W(z)}}{1+W(z)}
\ee
and 
\be \label{slow_roll-string-2}
\upeta = \frac{1}{\sqrt{\cR}}\Big(
\frac{1}{N} + \frac{W(z) (1-W(z))}{(1+W(z))^2}\cdot\frac{3 y^2}{\pi x^2}
\Big).
\ee
An interesting feature of \eq{slow_roll-string} and \eq{slow_roll-string-2} is
that, given $x$ and $y$, $\uve$ and $\upeta$ stay very small for a large range of
$N$ from the initial time, 
and then increase rapidly at around $N$ of the order of $N\sim x y^{-1} \ln y^{-1} $.
Typical plots of $\uve$ and $\upeta$ against $N$ are shown in the figures
\ref{e-s} and \ref{eta-s} 
\footnote{
The equation \eq{slow_roll-string-2} is divergent at $N=0$. This is due
to the fact that, for the interest of leading IR effect, 
we have only kept the leading  $N^2$ term in the
expression \eq{vtot-1} for $\d v$. For small $N$, one should also keep
the subleading order $N$ term and the resulting $\upeta$ is then regular. 
}

We emphasis
that the parameters $\uve$ and $\upeta$ introduced above are defined with
respect to the string frame Hubble parameter. 
In the next section, we
will go to the Einstein frame and show that time dependent effect
\eq{dRG-v} of the vev leads to an inflationary cosmology
whose slow roll parameters $\uve_E$, $\upeta_E$ are small and resembles
those of the standard slow roll inflationary model.

\section{Gravitational IR-Effect Driven Inflation}

\subsection{Einstein frame Hubble constant}

To examine the physical significance of the time dependent shift $\d
v$  of the vev
of the dilaton,  let us go to the Einstein frame. 
If we denote by $\vphi$ the dynamical part of $\phi$ above the vev, 
\be
\phi= v_{\rm eff} + \vphi,
\ee 
then in order to decouple the dilaton field $\vphi$ from  the
Hilbert-Einstein term, we may consider a  Weyl scaling
of the metric of the form
\be \label{weyl}
{g}_{\mu\nu}=g^E_{\mu\nu}e^{2(\frac{ \vphi}{\eta}  + \b) },
\ee
where $\b$ is an arbitrary function that is independent of the
dilaton field $\vphi$. 
Using \eq{weyl}, the scalar curvature $R$ in the string frame
can be written in terms of $g^E_{\mu\nu}$ and $\vphi$ as
\beq
R =
e^{-2( \frac{\vphi}{\eta}  + \b )}\left[R_E- 
6 g_E^{\mu\nu}\parl_{\mu} ( \frac{\vphi}{\eta}  + \b ) 
\parl_{\nu} ( \frac{\vphi}{\eta}  + \b ) -
6 \Box_E   ( \frac{\vphi}{\eta}  + \b )
\right], 
\label{RJE}
\eeq
and the action in the Einstein frame is 
\bea
S_{E}&=&\int\sqrt{-{ g_E}}d^4 x\left[\frac{M^2}{2}e^{2(\b -
    \frac{v_{\rm eff}}{\eta} )}
\Big(R_E - 6 g_E^{\mu\nu}\parl_{\mu} ( \frac{\vphi}{\eta}  + \b) 
\parl_{\nu} ( \frac{\vphi}{\eta}  + \b) - 6 \Box_E (
\frac{\vphi}{\eta}  + \b) \Big)
\right. \nn \\
&& \qquad \left. - V(\phi) e^{4 ( \frac{\vphi}{\eta}  + \b )}
-\frac{1}{2}e^{2 (\frac{\vphi}{\eta}  + \b )}
{g}_E^{\mu\nu} \del_\m (\vphi + v_{\rm eff})  \del_\n (\vphi + v_{\rm eff}) 
\right].
\label{acE}
\eea

Note that 
spatial dependence in $\b$ would lead the de Sitter solution \eq{H0a0}  
to an inhomogeneous background metric in the Einstein frame. As we are
interested in homogeneous metric as the cosmological description 
of the universe, so we will not consider spatial dependent $\b$. 
This gives the scale factor in the Einstein frame
\be \label{aE}
a_E (\t) = a_0 (\t) e^{-\frac{\d v}{\eta}  -\b}
\ee
where $a_0 (\t) $ is given by \eq{H0a0}.
We can also read off from \eq{acE} the Planck mass
\be \label{MP}
M_P = M e^{\b - \frac{v_{\rm eff}}{\eta}}.
\ee
To compute the Hubble parameter 
\be
H_E = 
\frac{1}{a_E} \frac{d a_E}{dt_E},
\ee 
we note that
\be \label{dtEdt}
dt_E =  e^{-(\b+ \frac{\d v}{\eta})} dt
\ee
as a result of the background Einstein metric
\be \label{ds-E}
ds_E^2 = e^{-2(\b+ \frac{\d v}{\eta})} (-dt^2 + a_0^2 d\bx^2).
\ee
We obtain 
\be \label{HE}
H_E = e^{\frac{\d v}{\eta}  + \b} (H_0 -\frac{d \b}{dt} -
\frac{1}{\eta} \frac{d \d v}{dt}).
\ee
In particular, we have
\be \label{HE1}
H_E = e^{\frac{\d v}{\eta}  + \b} H_0, 
\quad \mbox{if} \quad 
\dot{\b},\; \frac{1}{\eta}\d \dot{ v} \ll H_0.
\ee
As we will see later, this is the case of interest and relevance to us.

It is instructive to note that the relation \eq{HE1} may also be 
understood  using the Friedmann equation
\be
H_E^2 = \frac{U}{3M_P^2} 
\ee
where $U$ denotes the energy density for the 1-loop corrected 
vacuum $\phi = v_{\rm eff}$, i.e. $\vphi =0$. 
By putting $\vphi=0$ in \eq{acE}, we obtain 
\be \label{U-vac}
U = e^{4 \b}\Big[ V(v_{\rm eff}) + 3M^2 e^{-\frac{2v}{\eta}} 
\Big((\frac{d \b}{dt})^2 -\frac{1}{a_0^3 e^{-\frac{2\d v}{\eta}}} 
\frac{d}{dt} (a_0^3 e^{-\frac{2\d v}{\eta}}\frac{d \b}{dt}) \Big)
-\frac{1}{2} e^{\frac{2 \d v}{\eta}} (\frac{d \d v
    }{dt})^2 \Big]
\ee
where we have used the background Einstein metric \eq{ds-E}.
This gives the Hubble parameter
\be \label{HE2}
H_E^2 = e^{2(\b+ \frac{\d v}{\eta})} 
\big(H_0^2 + (\frac{d \b}{dt})^2 + \cdots \big),
\ee
where $\cdots$ denotes  contributions obtained from the third and fourth term
in \eq{U-vac}. The two terms listed above resemble those one would obtain from
\eq{HE}.  However the $\cdots$ terms are completely different. 
The reason for the discrepancy is simple: 
the expression \eq{U-vac} is not the correct vacuum energy density for
the state $\phi = v + \d v$ as it was obtained from the tree level
Einstein action without 
taking into full  account of the 1-loop IR
effects. Turning the argument around, we can use the Friedmann
equation to obtain the  vacuum energy density,
\be \label{U-vac1}
U =  e^{4 \b} (V(v) +  \d V) \Big(
 1 -\frac{1}{H_0}\frac{d \b}{dt} -
\frac{1}{H_0\eta} \frac{d \d v}{dt}
\Big)^2,
\ee
where $\d V$ denotes the 1-loop corrections, both UV and IR, to $V$.
As we discussed above, UV corrections are time independent. Typically,
$\d V$ receives a contribution of order $\Uplambda_{UV}^4$ from the zero point 
energy fluctuation. This term is much bigger than the other UV or IR corrections to
the potential. So the dominant time dependence comes from the
exponential prefactor and 
one recovers immediately \eq{HE1}.

A couple of remarks follows.
\ben
\item  So far $\b$ is general.  
While the relations \eq{MP}, \eq{HE1}
generally depend on $\b$, the ratio
\be \label{ratio}
\frac{H_E}{M_P} = \frac{H_0}{M e^{-v/\eta}} e^{2 \d v/\eta}
\ee
is independent of the choice of the frame and thus is an useful
quantity to ask physical question about it.

\item 
In the literature, IR effects of
the graviton loops have been studied rather extensively in various
setting and models.
The exponentiation of the IR
effects of the graviton loops  \eq{ratio} is new and is the 
main finding of this paper. Note  that $\d v$ is always negative 
for $\eta>0$ so we always get an exponential suppression/screening on the 
cosmological constant.
In general, the longer the elapsed time,
the greater will be the screening.

\item 
In the above analysis, we did not take into account of 
the quantum gravity corrections to the classical Einstein equation. 
This is a difficult problem since quantized gravity effect in de Sitter background is 
poorly understood in general. An embedding in string theory does not
help in this case since string theory in time dependent background has
met with a number of conceptual as well as technical difficulties. 
Due to a lack of  reliable mean to  compute these quantum corrections, we will  
ignore this issue in our current discussion. 

\een

\subsection{Slow roll inflation from gravitational IR-effect}

$M_P$ is time dependent for general $\b$.
Although it is an interesting 
scenario to consider a time dependent Newton constant, and it can
indeed be  easily
accommodated in our framework, 
however  given that it is not yet
universally accepted that such a variation does exist 
\footnote{
Current observational bound \cite{bdd} 
for the time-dependence of 
the Newton constant $8 \pi G=M_{P}^{-2}$ is small and is of the order of
$|{\dot G}/G| \sim 10^{-11} - 10^{-14}{\rm year}^{-1}$.
}, we will not consider this possibility in this paper.
Let us therefore consider the choice of frame with 
\be \label{ours-beta}
\b = \frac{\d v}{\eta}.
\ee
In this frame, 
we have a constant Planck scale
\be \label{MP1}
M_P = M e^{-v/\eta}.
\ee
This gives not just the natural UV cut off scale of the Einstein
frame action, but also the  UV cutoff scale $\Uplambda_{\rm UV}$
of the string frame action
\eq{action} in the presence of vev \eq{seq3}.
As it is easy to check that  
$\eta^{-1} d \d v/dt \ll H_0$ is satisfied,  
therefore the Hubble parameter acquires a time variation of the form 
\be \label{HE-ours}
H_E = H_0 e^{\frac{2 \d v}{\eta}} 
\ee
from the graviton IR loop effect.
In this frame, $H_0$ is also the initial value of the Einstein frame 
Hubble parameter. Note that generally one may consider a change of
frame by adding to \eq{ours-beta} an
arbitrary constant and still obtain a time independent Planck
mass. 
This would modify the definition \eq{MP1} of $M_P$ with a
multiplicative constant factor. However this is a physically equivalent frame
since  physics is unchanged if we express everything in terms of $M_P$.

As $\d v <0$, the Einstein frame metric describes an
inflationary cosmology with a Hubble parameter \eq{HE-ours} that
is decreasing in magnitude with time.  The slow roll parameters,
\be
\uve_E := \frac{ d}{d t_E} H_E^{-1}, \quad 
\upeta_E := \frac{1}{H_E  \uve_E} \frac{d\uve_E}{dt_E}
\ee
are given by
\be
\uve_E= 2 e^{\frac{\d v}{\eta}} \uve , 
\quad
\upeta_E=  e^{\frac{\d v}{\eta}}( \upeta - \uve)
\ee
for our model. We note that $N$, as defined in \eq{N-string} above, is
in fact equal to the number of e-folding $\cN_E$ in the
Einstein frame $d\cN_E := H_E dt_E$:
\be \label{NEN}
\cN_E = N.
\ee
That this is true independent of the choice of $\b$ can be seen
immediately from \eq{dtEdt} and \eq{HE1}. Due to the presence of the
suppression factor $  \exp {\frac{\d v}{\eta}} <1$, the Einstein 
slow roll parameters will be small as long as the $\uve, \upeta$ are
small. Since the later parameters control the backreaction to the
quantum field computation performed in the string frame, therefore as
long as we are in a regime of parameters \eq{limit-s} 
that we can trust the quantum 
field theory computations,
the corresponding cosmology in the Einstein
frame describes a slow roll inflation with
\be \label{limit-e}
\uve_E, \upeta_E \ll 1.
\ee

A couple of remarks are in order.
\ben

\item
In discussing the cosmological  consequences of any QFT calculation, it is
important to  employ an observable that is gauge invariant. 
In the above, 
we have used the time dependent IR effects of the vev of the
dilaton scalar field to deduce the existence of 
a screening effect on the cosmological constant.   
In a gauge
theory,  the quantum corrected effective potential  is generally
gauge dependent. While it does not necessarily mean 
that the vev of the scalar field,  
as can be determined equivalently from the effective potential,  is also gauge
dependent,  it is necessary to check whether this is the case. If
it is so,  it will be interesting to replace the vacuum expectation value
of the scalar field  with a gauge invariant order parameter and 
redo the analysis of this paper. Lessons learnt in a previous analysis
 \cite{bkreaction} may be useful.

\item  In the simplest model of inflation, inflationary expansion
  is driven by the slow rolling of an inflaton field down an
  almost flat  potential as in the slow roll inflation model. 
In our model,  inflationary expansion is driven by a different
mechanism, the IR
effects of the gravitons themselves. Slow roll inflation is achieved
without a slow roll potential. 

\item One of the obstacles in the slow roll model of inflation is that
  why is the inflaton mass so light. Expressed in terms of the eta
  parameter, it is required that
\be
\upeta_E = \frac{m_\phi^2}{3 H_E^2} \ll 1. 
\ee
Like the Higgs hierarchy problem, generically 
\be
m_\phi^2 \sim \Uplambda_{\rm UV}^2 \gg H_E^2.
\ee
Supersymmetry improves it a little since
contributions from bosons and fermions cancel precisely. However
supersymmetry is spontaneously broken during inflation and this leads
to an inflaton mass of order Hubble
\be
m_\phi^2 \sim H_E^2. 
\ee
and $\upeta$ is of  order one. The presence of large quantum
corrections to the eta parameter simply ruins the inflationary picture
predicted by the classical potential.

In our model, the dilaton field is sitting at the minimum of the quantum
corrected effective potential. Change of vacuum energy is not due to a
rolling of the dilaton field as in model with an inflaton, but
is due to the time dependent IR effect of graviton loops 
on the position of the minimum. It is all right for $m_\phi^2$ to 
receive large UV corrections, but these are time independent and
does not change the results of our model.  For example, the value of
$H_0$ will depend on the UV cutoff of the theory, but the time
dependence  in \eq{HE-ours} is not as it arises from IR quantum 
corrections. 
In order words, unlike the slow roll model, our model is free from 
the eta problem and we can trust the time evolution of the Hubble 
parameter.

\een

\subsection{Screening of the cosmological constant}

In the history of universe, there has been at least two
regimes of de Sitter phases, one is the inflation era in the early
universe, the other is the expansion of the current universe 
which is described by an de Sitter metric in the asymptotic future. 
For the inflationary phase, the  
inflation scale is constrained by the
Planck observation \cite{planck} of the amplitude of the CMB power spectrum to be 
\be
H_{E, {\rm inf}} \sim (\frac{r}{0.01})^{1/2} \times 10^{14} {\rm GeV},
\ee
where $r$ is the tensor-to-scalar ratio. For concreteness, let us
consider the case of $r \sim 0.01$ and so 
\be \label{H-inf}
H_{E, {\rm inf}} \sim 10^{14} {\rm GeV}.
\ee 
One of the interesting question about inflation is what set the scale
of inflation $H_{E, {\rm inf}}$? We can use \eq{HE-ours} to address
this. In the usual picture about quantum gravity, 
spacetime is highly quantized right after the big bang. 
After about one unit of Planck time, classical
geometry begins to make sense. 
Let us consider the situation where inflation started at about this time.
In this case, 
it is natural to take the
initial condition that the initial value of the Hubble parameter is
given by the Planck mass
\be \label{IC}
H_E(0) = M_P.
\ee
This corresponds to $x=1$. 
After expanding with a number $N$ of e-folding, the Hubble parameter is given by
\be
H_E(N) = M_P \;  \cR,
\ee
where $\cR$, as given by the \eq{RW}, determines the amount of
screening. 
We have thus found that the 
de Sitter symmetry breaking IR loop effects provide a screening of the
cosmological constant. Screening of the cosmological constant due to
IR effects of gravity has been  conjectured and argued for long ago.   
Our model provides a concrete set up where the 
screening mechanism and its effects can be calculated reliably. 

For a given model, $M, v$ and $\eta$ are given. Equivalently we can
use the Planck mass $M_P$, which set the scale, and 
the dimensionless parameters $x,y$ of
\eq{xy} to specify the model. 
As long as the parameters \eq{limit-s} are small, it is easy to
accommodate \eq{H-inf}  with an amount 
\be \label{RW4}
\cR =10^{-4}
\ee
of screening. 
In Table 1, we show for $x=1$ and different values of $y$, the 
values of $N$, $\uve$ and $|\upeta|$ giving 
$\cR =10^{-4}$.
\begin{table}[tbp]
\newcommand{\lw}[1]{\smash{\lower0.3ex\hbox{#1}}}
\begin{center}
\renewcommand{\arraystretch}{1.2}
\begin{tabular}{p{2cm}p{1.5cm}p{1.5cm}p{1.5cm}}
\toprule
 $y$&$N$&$\uve$&$|\upeta|$\\
\midrule
$1$ 
& 
$628$
&
$0.16$
&
$6\times 10^{-7}$
\\
$0.01$ 
&
$1000$
&
$0.22$
&
$0.06$
\\
$0.0001$
&
$77850$
&
$0.012$
&
$0.013$ \\
\bottomrule
\end{tabular}
\end{center}
\caption{The values of $N$, $\uve$ and $|\upeta|$ giving $\cR=10^{-4}$
for $x=1$ and different values of $y$.}
\label{table2}
\end{table}
In practice we want to achieve this amount
of screening before inflation ends. This is the point  where 
$\uve_E =1$.  One can play with the parameters $x, y$ and 
it is not hard to convince oneself that our $\uve_E$ can never exceed
1. What it means is
the IR effect of graviton loops on the vev is not sufficient to
end inflation;
and we need a new effect to change the behavior of the Hubble
parameter so that  $\frac{d}{dt_E} (a_E H_E)^{-1}$ change sign.
What could this effect be?

In the simplest slow roll inflation model, inflation ends when the
inflaton potential steepens (large $V'$) and the inflaton field picks up kinetic
energy. After inflation end, the inflaton starts to oscillate around
the global minimum of the potential and the energy of the inflaton
field is transferred to the standard model sector through a decay of
the inflaton field to the standard model particles. 
In the scenario described above,  
the universe
started to inflate right after or soon after big bang, with a Planck
scale Hubble parameter initially. The IR loop effect of the
graviton generates a screening on the Hubble parameter which resembles
the slow roll feature of the flat potential in slow roll inflation
model. 
In order to be able to transfer the energy stored in the dilaton to
the standard model matter fields, we assume that the dilaton is coupled to the 
standard model field $\psi$, for example
\be
\cL_{\rm int}  = \l \phi \psib \psi,
\ee 
Now $\l$
determines the rate of decay of the inflaton to the standard model
particles,  $\phi \to \psib \psi$, and this has a effect of decreasing
the Hubble parameter. This effect is usually small,  but in our model,
one can expect that the coupling $\l$ will also receive de Sitter
symmetry breaking IR corrections and becomes time dependent
\footnote{
 Similar
effects have been studied and reported in \cite{Kitamoto:2012ep,Kitamoto:2014gva}.
}.
It is possible that $\l(t)$ may become strong as time evolves and
effectively playing the role of a steepened potential and ends the
inflation.  This is an
interesting scenario and will be the subject of a separate paper.

As for the current universe, 
the current  value of the Hubble constant  $H_{E, {\rm now}}\simeq
10^{-42}{\rm GeV}$ is tiny. In terms of the Planck mass $M_P = 10^{18}
{\rm GeV}$, it is
\be
H_{E, {\rm now}} \simeq 10^{-60} M_P,
\ee
or in terms of the cosmological constant $\L_E = 3 H_E^2$:
\be
\L_{E,{\rm now}} \simeq 10^{-120} M_P^2.
\ee
The original cosmological constant problem is to understand why the
current cosmological constant is so much smaller than  $M_P^2$, the natural
value of the vacuum energy in a generic setting. Supersymmetry helps a
little but we still get a large hierarchy to explain. We will have
nothing to say about this problem except to say that in our model the Hubble
parameter is a function of time whose history is determined by the
dynamics of the theory and the initial condition, e.g. \eq{IC}.
A proper understanding of the process of
reheating is needed in order to understand what value the cosmology
constant take after reheating. This would serve as the initial
condition of the Hubble parameter which then evolve  to the small value
it takes nowadays.

\section{Conclusion and Discussion}

In this paper, we have considered an effective theory of gravity 
with which a dilaton field is coupled exponentially to. 
We found that the 1-loop IR
effects of the gravitons break the de Sitter symmetry of the
background, and constitute a 
time dependent contribution 
to the vev of
dilaton field. We note that, in the Einstein frame, this time
dependent effect is
exponentiated and acts to reduce the cosmological constant over
time. This provides a  
concrete mechanism of screening of the cosmological constant through
the IR effects of the graviton.

To determine the late time behaviour of the system,
we employ the DRG method to re-sum the leading IR logarithms. 
This allows us to follow the effects of the screening on
the cosmological evolution of the universe. 
In particular, we find
that one can have an inflation scenario driven entirely by the gravitational 
IR effect. This IR-driven inflation achieves 
all the standard features of the slow roll inflation
model such as having a slowing changing Hubble constant (small slow
roll parameters ) and sufficient amount of e-foldings etc. 
Moreover, since the UV divergence
of the theory is time independent and does not mix, at least in the
1-loop order, with the time dependent IR
effects; as a consequence, our model does not suffer from the eta
problem that 
baffles models with inflation driven by an inflaton potential.

To discuss the ending of inflation and reheating in our model, 
it is necessary to include in our model the coupling of the dilaton
field to matter fields. We speculated that the IR effect on the
dilaton-matter coupling may acts effectively like a damping term and 
provide a mechanism for the ending and reheating of the inflation. 

The dilaton-gravity sector of our model is specified by the Planck
mass scale and two dimensionless parameters, ratio of initial Hubble
constant with respect to the Planck mass, and the ratio of the
dilaton gravity coupling constant with respect to the Planck mass. It
is interesting to study the other signatures of the model,
such as the tensor scalar ratio $r$,  the primordial  non-Gaussanity
$f_{NL}$ and the tilt and other features of the power spectrum.

There is compelling evidence that the universe is presently
undergoing  a period of accelerated expansion
and the expansion is supported by some form of dark energy. 
However the nature of dark energy is mysterious and it is not known
whether it is a cosmological constant or
something else. If it is given by some form of 
quintessence, the kind of IR loop effects we studied in this paper
will exist and may play a role and leave signature
on the late time cosmology.

Our analysis is based on 1-loop UV and IR effects,
improved by a resumation of the leading IR logarithms. 
At higher loop orders, UV divergences may mix with the IR divergences
and leads to new effects. This however will depend  on the
UV completion of the effective theory. This is one way how Planckian
suppressed corrections may become relevant and how UV
sensitivity maybe regained in our model.


\vskip7mm
\section*{Acknowledgments}

We would like to thank Toshiaki Fujimori, Satoshi Iso, Hiroshi Isono,
Shoichi Kawamoto, Yoshihisa Kitazawa, Richard Woodard and 
Jackson Wu for valuable discussions. 
This work is
supported in part by  the National Center of Theoretical Science
(NCTS) and the 
grant  101-2112-M-007-021-MY3 of the Ministry of Science and
Technology of Taiwan.

\appendix

\section{Interaction Terms of Gravitons and the Dilaton}
\label{app-int}

In this appendix, we give the interaction terms of the graviton
and the dilaton. In this article, we need only the
three-point vertices, $hh{\hat \phi}$ and ${\hat \omega}{\hat
\omega}{\hat \phi}$, to calculate the 1-loop tadpole diagrams generated by
the graviton loop. For completeness, we list also the
interaction terms up to cubic order in the graviton
perturbations, $h_{\mu\nu}$ and $\omega$, and quintic order in the
scalar perturbation ${\tilde \phi}$,
which may be useful for other applications of our model. 

Our total interaction Lagrangian 
is given by 
\be
\cL_{\rm int}=\cL_1+\cL_2+\cL_3,
\ee
where
\beq
\cL_1&=&\frac{M^2}{2}\sqrt{-g}e^{-\phi/\eta}R,\\
\cL_2&=&-\frac{1}{2}\sqrt{-g}g^{\mu\nu}\partial_{\mu}\phi\partial_{\nu}\phi,\\
\cL_3&=&-\sqrt{-g}V(\phi).
\eeq 
Presented  according to the
order of the order of graviton fluctuations, we have: 

\noindent (1) Zeroth order in $h$ and $\omega$:
\beq
\frac{8\kappa^2}{H_0^2a_0^4}\cL_1^{(0)}&=&
 -\frac{2}{\eta^3}\zeta^3{\hat \phi}^3
+ \frac{1}{\eta^4}\zeta^4{\hat \phi}^4 ,\\
\cL_2^{(0)}&=&0,\\
\frac{1}{a_0^4}\cL_3^{(0)}&=&-
 \frac{\xi}{6}\zeta^3{\hat \phi}^3
-\frac{\lambda}{24}\zeta^4{\hat \phi}^4 ,
\eeq
where
\be
\xi:= V^{(3)}(v),\quad \lambda:=V^{(4)}(v)
\ee
and we have used
\be
{\tilde \phi}=\zeta{\hat \phi}.
\ee

\noindent (2) Linear order in $h$ and $\omega$:
\beq
\frac{\kappa}{2a_0^2}\cL_1^{(1)}&=&
\left[\parl_{\r}\parl_{\nu}h^{\r\nu}-\frac{6}{\tau}\parl_{\r}h^{\r0}
+\frac{12}{\tau^2}h^{00}-6\parl_{\mu}\parl^{\mu}{\hat \omega}
-\frac{12}{\tau}\parl_0{\hat \omega}+\frac{24}{\tau^2}{\hat \omega} 
\right.\notag\\
&&\label{int11}
 \left. -\frac{6\zeta}{\kappa\eta}
\left(\parl_{\mu}\parl^{\mu}{\hat \phi}+\frac{2}{\tau}\parl_0{\hat
  \phi}
-\frac{4}{\tau^2}{\hat \phi}\right)\right]
\left(\frac{\zeta^2}{\eta^2}{\hat \phi}^2
-\frac{2}{3}\frac{\zeta^3}{\eta^3}{\hat \phi}^3+\frac{1}{3}
\frac{\zeta^4}{\eta^4}{\hat \phi}^4\right),\\
\label{int21}
-\frac{2}{ \kappa \zeta^2 a_0^2}\cL_2^{(1)}&=&\parl_{\mu}{\hat \phi}
\parl_{\nu}{\hat \phi}\left(2{\hat \omega}\eta^{\mu\nu}
+\frac{2\zeta}{\kappa\eta}{\hat \phi}\eta^{\mu\nu}-h^{\mu\nu}\right),\\
\label{int31}
-\frac{1}{4\kappa a_0^4}\cL_3^{(1)}&=&\left({\hat \omega}
+\frac{\zeta}{\kappa\eta}{\hat \phi}\right)
\left(\frac{\sigma^2}{2} \zeta^2{\hat \phi}^2+\frac{ \xi}{6}\zeta^3 {\hat \phi}^3
+\frac{\lambda}{24}\zeta^4{\hat \phi}^4\right).
\eeq

\noindent  (3) Quadratic order in $h$ and $\omega$.
\beq
\frac{1}{a_0^2}\cL_1^{(2)}&=&
\left[
 \uline{-\frac14\parl_{\mu}h_{\r\sig}\parl^{\mu}h^{\r\sig}-\frac12
\parl_{\mu}(h^{\mu}_{\ \r}\parl_{\nu}h^{\r\nu})-\frac12h^{\mu}_{\ \r}
\parl_{\mu}\parl_{\nu}h^{\r\nu}+2{\hat \omega}\parl_{\mu}
\parl_{\nu}h^{\mu\nu}}\right.  \notag\\
&&  \uline{+6\parl_{\mu}(h^{\mu\nu}\parl_{\nu}{\hat \omega})- 
\frac{12}{\tau}\parl_{\mu}(h^{0\mu}{\hat \omega})+\frac{3}{\tau}\
\parl_{\mu}(h^{0\nu}h_{\nu}^{\ \mu})-\frac{6}{\tau^2}h^{\r0}h_{\r}^{\ 0}
+\frac{24}{\tau^2}h^{00}{\hat \omega}}\notag\\
&&  \uline{ 
-6\parl_{\mu}({\hat \omega}\parl^{\mu}{\hat \omega})
-6{\hat \omega}\parl_{\mu}\parl^{\mu}{\hat \omega}
- \frac{24}{\tau}{\hat \omega}\parl_0{\hat \omega}
+\frac{24}{\tau^2}{\hat \omega}^2}
\notag\\
&&
-\frac{6\zeta}{\kappa\eta}
\left(-\frac13{\hat \phi}\parl_{\mu}\parl_{\nu}h^{\mu\nu}
-\parl_{\mu}(h^{\mu\nu}\parl_{\nu}{\hat \phi})
+\frac{2}{\tau}\parl_{\mu}(h^{0\mu}{\hat \phi})
-\frac{4}{\tau^2}h^{00}{\hat \phi}
+{\hat \phi}\parl_{\mu}\parl^{\mu}{\hat \omega} 
\right.\notag\\
&&\quad\qquad\left.
+{\hat \omega}\parl_{\mu}\parl^{\mu}{\hat \phi}
+\parl_{\mu}({\hat \phi}\parl^{\mu}{\hat \omega}
+{\hat \omega}\parl^{\mu}{\hat \phi})
+\frac{4}{\tau}{\hat \phi}\parl_0{\hat \omega}
+\frac{4}{\tau}{\hat \omega}\parl_0{\hat \phi}
-\frac{8}{\tau^2}{\hat \omega}{\hat \phi}
\right)\notag\\
&& 
\left.-\frac{6\zeta^2}{\kappa^2\eta^2}
\left(\parl_{\mu}({\hat \phi}\parl^{\mu}{\hat \phi})
+{\hat \phi}\parl_{\mu}\parl^{\mu}{\hat \phi}
+\frac{4}{\tau}{\hat \phi}\parl_0{\hat \phi}
-\frac{4}{\tau^2}{\hat \phi}^2
\right)\right]\notag\\
&&\label{int12}
\times
\left(\uline{-\frac{2\zeta}{\eta}{\hat \phi}}
+\frac{\zeta^2}{\eta^2}{\hat \phi}^2
-\frac{2\zeta^3}{3\eta^3}{\hat \phi}^3
+\frac{\zeta^4}{3\eta^4}{\hat \phi}^4\right), \\
-\frac{4}{ \kappa^2 \zeta^2 a_0^2}\cL_2^{(2)}&=&
\parl_{\mu}{\hat \phi}\parl_{\nu}{\hat \phi}
\left[h^{\mu\r}h_{\r}^{\ \nu}
+4\eta^{\mu\nu}{\hat \omega}^2
-4h^{\mu\nu}{\hat \omega}
+\frac{4\zeta}{\kappa\eta}\Big(2\eta^{\mu\nu}{\hat \omega}{\hat \phi}
-h^{\mu\nu}{\hat \phi}+\frac{\zeta}{\kappa\eta}\eta^{\mu\nu}{\hat
  \phi}^2\Big)
\right],\notag\\
\label{int22}
\\
\label{int32}
-\frac{1}{8\kappa^2 a_0^4}\cL_3^{(2)}&=&
 \Big( \uline{{\hat \omega}^2}
 +\frac{2\zeta}{\kappa\eta}{\hat \omega}{\hat \phi}
 +\frac{\zeta^2}{\kappa^2\eta^2}{\hat \phi}^2\Big)
 \Big(\uline{\r\zeta{\hat \phi}}
+ \frac{\sigma^2}{2} \zeta^2{\hat \phi}^2+ \frac{\xi}{6} \zeta^3{\hat \phi}^3
+\frac{\lambda}{24}\zeta^4{\hat \phi}^4\Big).
\eeq
In deriving \eqref{int11} -- \eqref{int32} we have used \eqref{canonical1}.
The underlined terms in  the above formula 
correspond to the vertices of the one-loop tadpole 
diagrams investigated in section 2.3.

\section{Comments on the IR Divergence of Two-Point Function}

\subsection{de Sitter space}

It is instructive to comment on the 
origin of time dependent de Sitter symmetry breaking
term in the propagator of a massless minimally coupled
scalar field. 
The mode expansion  for  a massless
minimally coupled scalar in de Sitter background in the Bunch-Davies vacuum  
is given by
\be
\chi =\frac{1}{(2\pi)^{3/2} a_0(\t)} \int d^3 p \; [a_{\bp}e^{i
    {\bp}\cdot{\bx}}\chi_p(\t)+{\rm h.c.} ] ,\quad
\chi_p (\tau) = \frac{1}{\sqrt{2p}}\left(1- \frac{i}{p\t} \right) e^{-ip\t}.
\ee
From this we obtain the propagator
\be
\langle\chi(x)\chi(x')\rangle
=\int \frac{d^3 p}{(2\pi)^3} \frac{H_0^2 \t \t'}{2p}\left(1-\frac{i}{p\t}\right) 
\left(1+\frac{i}{p\t'}\right)  e^{-ip(\t-\t')+ i  \bp\cdot (\bx-\bx')}.
\ee
The propagator has UV as well as IR divergences. The UV divergence
reflects the fact that the considered Lagrangian is a valid effective
description only down to a certain physical distance scale $\ell$ 
and the divergence can be regulated by imposing a  cutoff
\be \label{UV-reg}
P \leq \Uplambda_{\rm UV} := \frac{1}{\ell}
\ee
on the physical momentum
 $P := p /a_0(t)$.
The IR divergence
\be\label{IR-div}
\langle\chi(x)\chi(x')\rangle_{\rm IR} \sim
\int_0 \frac{d^3 P}{(2\pi)^3}\frac{H_0^2 }{2P^3},
\ee
on the other hand, is due to legitimate physical effects occurring at very long
distances. In the present case of a de Sitter background,
assume that initially the universe has a physical size 
$L_0$ where the different parts of the universe were in causal
contact, then  a sensible IR regulator is to put the universe in a
box of size 
$L(t)= a_0(t) L_0 $. This cuts out contributions from distance
scale larger than those that can be related by causal effects.
In terms of physical momentum, this corresponds to a cutoff
\be \label{IR-reg}
P\geq P_{\rm min} := \frac{1}{a_0(t) L_0}.
\ee
We emphasize that, unlike the IR regulator whose time dependence is
dedicated by the associated physics, the UV regulator \eq{UV-reg} is 
independent of time, and so UV divergences are taken care of by
the standard UV renormalization techniques. In contrast,  IR
divergences give rise to time growing de Sitter symmetry breaking
effect in the theory. The same time dependent factor also arises in the
graviton propagator.

\subsection{ Power law Friedmann-Robertson-Walker metric}

The result \eq{IR-div} about 
IR divergence can be easily generalized to the more 
general case of a spatially flat Friedmann-Robertson-Walker metric
\be
ds^2 = -dt^2 + a_0^2(t) d\bx^2,
\ee  
with a  power-law scale factor \cite{Ford:1977in,Allen:1986dd}
\be \label{power-a}
a_0(t) = w t^c \quad \mbox{where $w$ and $c$ are constants}.
\ee
Note that $c=\frac{1}{2}$ corresponds to radiation dominated era, 
$c=\frac{2}{3}$ corresponds to matter dominated era,
and $c\to \infty$ corresponds to de Sitter space \cite{Ford:1977in}.
A minimally coupled massless scalar in this metric has the 
equation of motion 
\be
-a_0^{-3} \del_t (a_0^3 \del_t \chi) + a_0^{-2} \del_i^2 \chi =0.
\ee
Let us introduce the conformal time $\t$ defined by (for $c \neq 1$),
\beq
\t :=\int^t \frac{dt'}{a_0(t')}
=\frac{1}{(1-c) w}t^{1-c}. \label{taud}
\eeq
In terms of $\t$, the scale factor can be written as
\beq
a_0= k \t^{\frac{c}{1-c}},\quad k =w [w(1-c)]^{\frac{c}{1-c}}
\eeq
and the field equation becomes
\beq
\del_\t (a_0^2 \del_\t) \chi-a_0^2\partial_i^2\chi=0.
\eeq
The mode function $\chi_p$ of $\chi$: 
\be
\chi =\frac{1}{(2\pi)^{3/2} a_0(\t)} \int d^3 p \; [a_{\bp}e^{i
    {\bp}\cdot{\bx}}\chi_p(\t)+{\rm h.c.} ] 
\ee
can be solved exactly in terms of the Hankel functions, 
\beq
\chi_p=c_1(p)\; \t^{\frac{1}{2}}H^{(1)}_{\nu} (p \t)
+c_2(p) \t^{\frac{1}{2}} H^{(2)}_{\nu} (p \t)
\label{sol}
\eeq
where 
\beq
b :=\frac{1-c}{1-3c},\quad \nu :=\frac{1}{|2b|} >0.
\eeq
Canonical quantization constraints the coefficients $c_1$ and $c_2$ 
to satisfy the normalization condition
\be
\chi_p \frac{d \chi_p^* }{d \t} - 
\chi_p^* \frac{d \chi_p }{d \t} = i.
\ee
This gives
\be
|c_2|^2-|c_1|^2=\frac{\pi}{4}\label{cr}
\ee
A convenient parametrization of the solution is 
\be
c_1=\sqrt{\frac{\pi}{4}}\sinh \alpha e^{i\beta},\quad  
c_2=\sqrt{\frac{\pi}{4}}\cosh \alpha 
\ee 
for real 
$\alpha$, $\beta$, where we have, for convenience, taken 
$c_2$ to be real and positive. 
In general an overall phase factor for $c_1$ and
$c_2$ can be inserted but it does not show up 
in any physical quantity. Thus it is sufficient to consider the two
parameters 
family ($\alpha$,$\beta$). In the de Sitter case, 
the Bunch-Davis vacuum corresponds to the choice 
$\alpha=0$. The
$\alpha$-vacua is  parametrized by ($\alpha$, $\beta$), $\a \neq 0$.

For quantum field theory in curved spacetime, it is customary to 
impose the Hadamard condition which states
that the short distance singularity structure of the
two point function should be closed to that for the Minkowski space 
\be
 \chi_p  \sim \frac{1}{\sqrt{2p}}e^{-ip \tau}, \quad p \t \gg 1.
\ee
Using the asymptotic expansion of the Hankel function for large $z \gg
1$, 
\be
H^{(1)}_{\nu}(z)\sim\sqrt{\frac{2}{\pi z}}e^{i(z-\pi\nu/2-\pi/4)},\quad
H^{(2)}_{\nu}(z)\sim\sqrt{\frac{2}{\pi z}}e^{-i(z-\pi\nu/2-\pi/4)}.
\ee
we find that the mode function behaves in the high energies limit as
\be
a_0 \chi_p\sim \frac{1}{\sqrt{2p}} (
c_1 e^{ i p \t }+c_2 e^{- i p \t}) ,
\ee
This amounts to the choice of the coefficients:
\be  \label{c1c2}
c_1=0, \quad |c_2|^2 = \frac{\pi}{4}.
\ee

We are interested in the IR behaviour of the two point function
\be \label{2pt}
\langle\chi(x)\chi(x') \rangle= \frac{1}{(2\pi)^3 a_0(\t) a_0(\t')} 
\int d^3p 
e^{i{\bp}\cdot({\bx}-{\bx}')}\chi_p(\tau)\chi_p^*(\tau').
\ee
For small $p$, we use the asymptotic behavior of the Hankel functions
(for $\Re \n>0$),
\be
H_\n^{(1)}(z) \sim - H_\n^{(2)} (z) \sim -\frac{i}{\pi} \G(\n)
(z/2)^{-\n},\quad z \to 0, 
\ee
then, apart from a constant factor, the small $p$ behavior of the 
two point function is given by
\be
\langle\chi(x)\chi(x') \rangle = \mbox{const.} \times
\int d^3 p\frac{|c_1-c_2|^2}{p^{2\n}} \times \frac{1}{(\t\t')^\b},
\ee
where 
\be
\b : = \n - (2b)^{-1} = 
\begin{cases}
2\n, & \mbox{for $b<0$, \; i.e. $1/3 <c<1$},   \\
0, & \mbox{for $b>0$,  \; i.e. $c< 1/3$ or $c>1$}. 
\end{cases}
\ee
We will adopt the Hadamard condition and so \eq{c1c2} implies that
the two point function  acquires an IR divergence from
the momentum integration if $| (1-3c)/(1-c)| \geq 3$, i.e.
\be
\frac{2}{3} \leq c \leq 1 \quad \mbox{or} \quad c >1.
\ee 
Thus apart from the de Sitter space ($c\to \infty$), 
the scalar two point function 
is also IR divergent  in the
matter dominated era $c=2/3$ and hence picks up the same time growing 
logarithmic factor $\ln (a_0(\t) a_0(\t'))$ after an IR cutoff is introduced
\footnote{
The two point function is IR finite without cutoff
in the radiation dominated era $c=1/2$.}.
However there is an important difference between the two cases: there
is an additional time dependent factor $(\t \t')^{-3} 
= a_0(\t)^{-3/2} a_0(\t')^{-3/2}$ in the case of matter
dominated era.
This factor actually decreases to zero faster than the growth
of the factor $\ln (a_0(\t) a_0(\t'))$, therefore we expect that the
graviton loop in the matter dominated era does not induces any screening
effect in late times.

\section{Light Field Condition for the Dilaton} 
\label{app-m2}

In this appendix we investigate the parameter region from the
constraint on 
the dilaton mass $m_{\hat \phi}\ll H^2$ which is needed for the
approximation  of the
massive scalar propagator \eqref{smpropagator3}. Let us recall  
\bea
m_{\hat \phi}^2 =\zeta^2\left(\sig^2+\frac{4\r}{\eta}\right)
,\quad
H_0^2=\frac{1}{3M^2}e^{2v/\eta}V_0,\quad \rho=-\frac{4V_0}{\eta}\label{He}
\eea
The condition is written as 
\bea
\frac{m_{\hat \phi}^2}{H_0^2}=\zeta^2\left(\frac{\sigma^2}{H_0^2}-
\frac{48M^2}{\eta^2}e^{-2v/\eta}\right)\ll 1.\label{ratio1}
\eea
For the two terms in the parenthesis to be small separately , we require
\be
\frac{16 V_0}{\eta^2}<\s^2\ll \frac{1}{3M^2}e^{2v/\eta}V_0, \qquad
\eta^2 \gg 48M^2e^{-2v/\eta}, \label{smallmass}
\ee 
where the lower bound of $\sigma$ comes from \eqref{sigmastable}. 
At the same time, we have $\zeta \simeq 1$.
We find that as long as the second constraint in \eqref{smallmass} 
is satisfied, there will be a broad range for  $\sigma^2$ and
\eqref{ratio1} 
can be easily satisfied. In terms of the parameter $y$ introduced in
\eq{xy}, the  second constraint in \eqref{smallmass} reads
\be
y^2 \ll \frac{1}{48}. 
\ee
In a different way, \eqref{ratio1} can be satisfied by requiring a 
balanced cancellation of two terms in the parenthesis in
\eqref{ratio1}. This 
leads to $\sigma\sim 16V_0/\eta^2$ and will need some fine tuning in
this case.


\end{document}